\documentclass[useAMS]{mn2e}

\usepackage{graphicx}

\title[Hen~2-39]{SALT reveals the barium central star of the planetary nebula Hen~2-39\thanks{Based on observations made with the Southern African Large Telescope (SALT) under programmes 2011-3-RSA-029 and 2012-1-RSA-009, the Very Large Telescope (VLT) at Paranal Observatory under programme 088.D-0750(A), the South African Astronomical Observatory (SAAO) 1.0-m and 1.9-m telescopes, and the New Technology Telescope (NTT) at La Silla Observatory under programme 090.D-0693(A).}}

\author[B. Miszalski et al.]{B. Miszalski,$^{1,2}$\thanks{E-mail: brent@saao.ac.za} H.~M.~J. Boffin,$^{3}$  D. Jones,$^{3}$ A. I. Karakas,$^{4}$ J. K\"oppen,$^{5}$\newauthor A. A. Tyndall,$^{3,6}$ S. S. Mohamed,$^{1}$ P. Rodr\'iguez-Gil$^{7,8,9}$ and M. Santander-Garc\'ia$^{10,11}$\\
$^{1}$South African Astronomical Observatory, PO Box 9, Observatory, 7935, South Africa\\
$^{2}$Southern African Large Telescope Foundation, PO Box 9, Observatory, 7935, South Africa\\
$^{3}$European Southern Observatory, Alonso de Cordova 3107, Casilla 19001, Santiago, Chile\\
$^{4}$Research School of Astronomy and Astrophysics, Australian National University, Canberra, ACT 2611, Australia\\
$^{5}$Observatoire astronomique de Strasbourg, Universit\'e de Strasbourg, CNRS, UMR 7550, 11 rue de l'Universit\'e, F-67000 Strasbourg, France\\
$^{6}$Jodrell Bank Centre for Astrophysics, School of Physics and Astronomy, University of Manchester, M13 9PL, UK\\
$^{7}$Departamento de Astrof{\'{\i}}sica, Universidad de La Laguna, La Laguna, E-38204, Santa Cruz de Tenerife, Spain\\
$^{8}$Instituto de Astrof{\'{\i}}sica de Canarias, V\'ia L\'actea, s/n, La Laguna, E-38205, Santa Cruz de Tenerife, Spain\\
$^{9}$Isaac Newton Group of Telescopes, Apartado de Correos 321, Santa Cruz de La Palma, E-38700, Spain\\
$^{10}$Observatorio Astron\'omico Nacional, Ap 112, 28803 Alcal\'a de Henares, Spain\\
$^{11}$CAB, INTA-CSIC, Ctra de Torrej\'on a Ajalvir, km 4, 28850 Torrej\'on de Ardoz, Madrid, Spain
}
\begin{document}

\date{Accepted Received ; in original form }

\pagerange{\pageref{firstpage}--\pageref{lastpage}} \pubyear{2012}

\maketitle

\label{firstpage}

\begin{abstract}
   Classical barium stars are binary systems which consist of a late-type giant enriched in carbon and slow neutron capture (s-process) elements and an evolved white dwarf (WD) that is invisible at optical wavelengths. The youngest observed barium stars are surrounded by planetary nebulae (PNe), ejected soon after the wind accretion of polluted material when the WD was in its preceeding asymptotic giant branch (AGB) phase. Such systems are rare but powerful laboratories for studying AGB nucleosynthesis as we can measure the chemical abundances of both the polluted star and the nebula ejected by the polluter. Here we present evidence for a barium star in the PN Hen~2-39 (PN G283.8$-$04.2) as one of only a few known systems. The polluted giant is very similar to that found in WeBo~1 (PN G135.6$+$01.0). It is a cool ($T_\mathrm{eff}=4250\pm150$ K) giant enhanced in carbon ([C/H]=$0.42\pm0.02$ dex) and barium ([Ba/Fe]=$1.50\pm0.25$ dex). A spectral type of C-R3 C$_2$4 nominally places Hen~2-39 amongst the peculiar early R-type carbon stars, however the barium enhancement and likely binary status mean that it is more likely to be a barium star with similar properties, rather than a true member of this class. An AGB star model of initial mass 1.8 $M_\odot$ and a relatively large carbon pocket size can reproduce the observed abundances well, provided mass is transferred in a highly conservative way from the AGB star to the polluted star (e.g. wind Roche-lobe overflow). It also shows signs of chromospheric activity and photometric variability with a possible rotation period of $\sim$5.5 days likely induced by wind accretion. The nebula exhibits an apparent ring morphology in keeping with the other PNe around barium stars (WeBo~1 and A~70) and shows a high degree of ionization implying the presence of an invisible hot pre-WD companion that will require confirmation with UV observations. In contrast to A~70, the nebular chemical abundance pattern is consistent with non-Type I PNe, in keeping with the trend found from nebular s-process studies that non-Type I PNe are more likely to be s-process enhanced. 

\end{abstract}


   \section{Introduction}

   Only a handful of central stars of planetary nebulae (CSPN) have been shown to be binaries in which the dominant component at optical wavelengths is a sub-giant or giant star enriched in carbon and slow neutron capture process (s-process) elements (Bond, Ciardullo \& Meakes 1993; Th\'evenin \& Jasniewicz 1997; Bond, Pollacco \& Webbink 2003; Miszalski et al. 2012; Siegel et al. 2012). 
   These binary CSPN are the progenitors of barium stars whose standard formation scenario (Bidelman \& Keenan 1951) involves the companion accreting a carbon and s-process enhanced wind (Boffin \& Jorissen 1988) from the pre-WD progenitor when it experiences thermal pulses during its asymptotic giant branch (AGB) phase (Busso et al. 1999; Herwig 2005). Those barium stars with the shortest orbital periods may also have experienced mass transfer via Roche-Lobe overflow (e.g. Han et al. 1995) or wind Roche-Lobe overflow (e.g. Mohamed \& Podsiadlowski 2007, 2011; Abate et al. 2013). McClure, Fletcher \& Nemec (1980) found all barium stars to have WD companions, however directly observing them is difficult against the glare of the typically luminous giant secondary even at ultraviolet (UV) wavelengths. Recent direct UV detections include the WD companions of main-sequence barium stars (Gray et al. 2011) and the evolved pre-WD components of barium CSPN (Miszalski et al. 2012; Siegel et al. 2012). There is now no doubt that barium stars form via binary interactions, although the real challenge at present lies in reproducing the orbital parameters (period and eccentricity) using population synthesis models (e.g. Izzard et al. 2010).
   
   Most barium CSPN will have orbital periods of several hundred days or longer, in line with other barium stars (Jorissen et al. 1998), but none have yet been measured for the small sample known. An orbital period of 1.16 d was, however, measured in the post-common-envelope nucleus of PN G054.2$-$03.4 (The Necklace, Corradi et al. 2011), notable for its unique carbon dwarf secondary (C/O$>$1; Miszalski, Boffin \& Corradi 2013). In this case the mass transfer process may have also produced an accretion disk from which the observed collimated outflows or jets were launched. 

   The short time spent during the PN phase ($\sim10^4$ years) makes barium CSPN a potentially very powerful platform for studying AGB nucleosynthesis in that we simultaneously see both the polluted s-process rich cool star and the nebula ejected by the polluting star. In recent years nebular s-process abundance measurements have also become possible (e.g. Sterling \& Dinerstein 2008). When combined with traditional nebular abundance analysis, they too will allow for new insights into AGB nucleosynthesis models (Karakas \& Lugaro 2010). Sterling \& Dinerstein (2008) and Karakas et al. (2009) found a tendency for Type I PNe (Peimbert \& Torres-Peimbert 1983; Kingsburgh \& Barlow 1994), i.e. those PNe that show high N/O and He/H abundance ratios traditionally thought to originate from intermediate-mass AGB stars ($M\ga2.5$ $M_\odot$, Kingsburgh \& Barlow 1994), to be less s-process enhanced than non-Type I PNe. Barium CSPN offer a complementary or even preferred platform for this work, with stellar abundance analysis of the cool star potentially providing both low and high s-process elements (e.g. Abia et al. 2002) and a direct measurement of the metallicity via [Fe/H] rather than potentially problematic nebular metallicity indicators (e.g. [O/H], see Karakas et al. 2009). It is therefore essential to find new examples of these rare objects to realise their unique capabilities. 
   
   As part of an ongoing campaign to study relatively faint CSPN with 8--10 m class telescopes we have continued with the strategy outlined in Miszalski et al. (2012), namely to identify similar binaries by selecting for near-infrared (NIR) colours that suggest the presence of a cool star. One object that stood out in the $J-H$ and $H-K_s$ colour-colour space of the Two Micron All Sky Survey (2MASS; Skrutskie et al. 2006) was the previously unstudied central star of Hen~2-39 (PN G283.8$-$04.2, Henize 1967). The relatively bright central star ($J=13.47$ mag) has NIR colours ($J-H=0.86$ mag and $H-K_s=0.28$ mag) that resemble an unreddened M-giant. 
   
   In this paper we report on observations of Hen~2-39 that reveal its barium star nucleus and its peculiar nebular morphology. Section \ref{sec:obs} describes the observations taken of Hen~2-39. Sections \ref{sec:cspn} and \ref{sec:neb} analyse the observations of the central star and the nebula, respectively. The distance to Hen~2-39 is discussed in Sect. \ref{sec:distance}, while Sect. \ref{sec:agb} and \ref{sec:agbresults} describe AGB models as a plausible explanation for Hen~2-39, followed by a discussion of the stellar and nebular chemical abundances in Sect. \ref{sec:comb}. We conclude in Sect. \ref{sec:conclusion}.

   \section{Observations}
   \label{sec:obs}
   \subsection{Spectroscopy}
   We obtained spectroscopic observations of Hen~2-39 with the Southern African Large Telescope (SALT; Buckley, Swart \& Meiring 2006; O'Donoghue et al. 2006) under programmes 2011-3-RSA-029 and 2012-1-RSA-009. Table \ref{tab:obs} outlines the observations made with the Robert Stobie Spectrograph (RSS; Burgh et al. 2003; Kobulnicky et al. 2003) where the position angle (PA) of the slit was always 0$^\circ$. Basic reductions were applied using the PySALT\footnote{http://pysalt.salt.ac.za} package (Crawford et al. 2010). Contemporaneous spectroscopic flat fields were taken with the first two RSS spectra and were incorporated in the PySALT reductions. Cosmic ray events were cleaned using the \textsc{lacosmic} package (van Dokkum 2001). Wavelength calibration of arc lamp exposures was performed using standard \textsc{IRAF} tasks \textsc{identify}, \textsc{reidentify}, \textsc{fitcoords} and \textsc{transform} by identfying the arc lines in each row and applying a geometric transformation to the data frames.  
   
   The extraction of one-dimensional spectra was complicated by the highly non-uniform stellar continuum and the variable spatial profile of nebular emission lines due to the apparent ring shape of the nebula (see Sect. \ref{sec:img}). Two extractions were made from the first RSS spectrum taken with the PG2300 grating, one covering the whole nebula and central star, and the other only the central star, with both having sky background subtracted from outside the nebula. The weakness of the stellar continuum in this spectrum for $\lambda\la4300$ \AA\ introduced minimal contamination into emission line intensities (see Sect. \ref{sec:neb}). For the spectrum taken with the PG900 grating, similar extractions were made as for the first PG2300 spectrum, except this time a scaled version of the stellar spectrum, with nebular lines interpolated over, was subtracted from the nebular spectrum. This produced a nebular spectrum with a uniform continuum from which to measure emission line intensities (see Sect. \ref{sec:neb}). This cleaned nebular spectrum was also rescaled and then subtracted from the CSPN spectrum to isolate the CSPN spectrum with only large subtraction residuals on the brightest lines (that were also interpolated over). For the second PG2300 spectrum only the CSPN was extracted with background sky subtraction and no nebular subtraction was made. The spatial centre of the nebular emission lines were fit for a reliable nebular radial velocity (RV) measurement with its contemporaneous thorium-argon arc lamp exposure (see later).

\begin{figure*}
   \begin{center}
      \includegraphics[scale=0.8,angle=270]{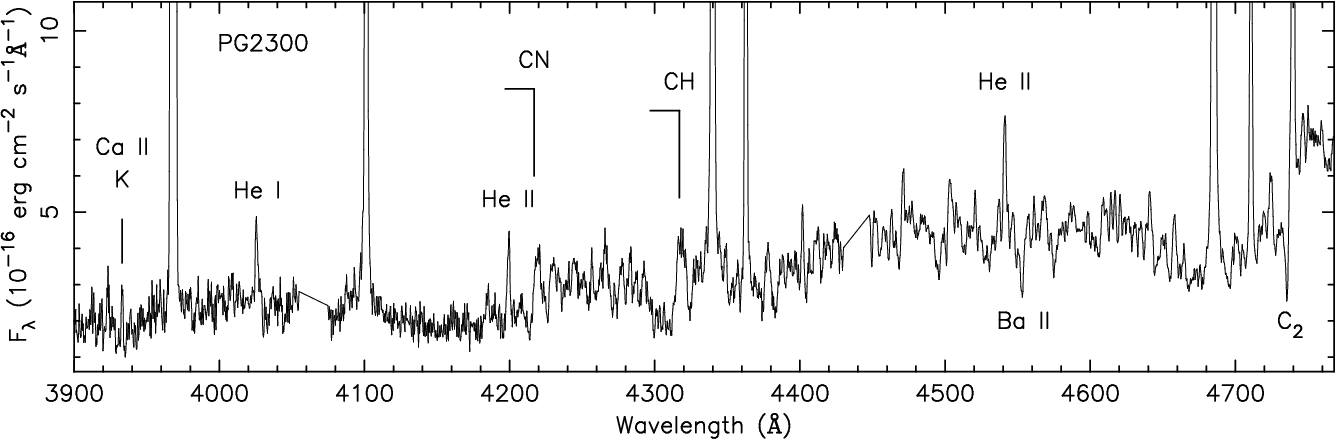}
      \includegraphics[scale=0.8,angle=270]{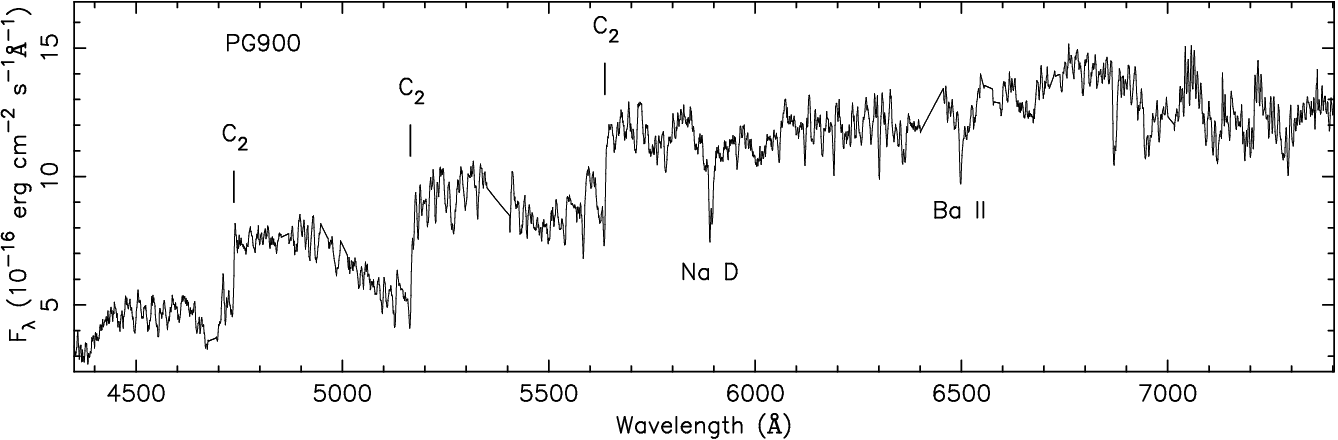}
      \includegraphics[scale=0.8,angle=270]{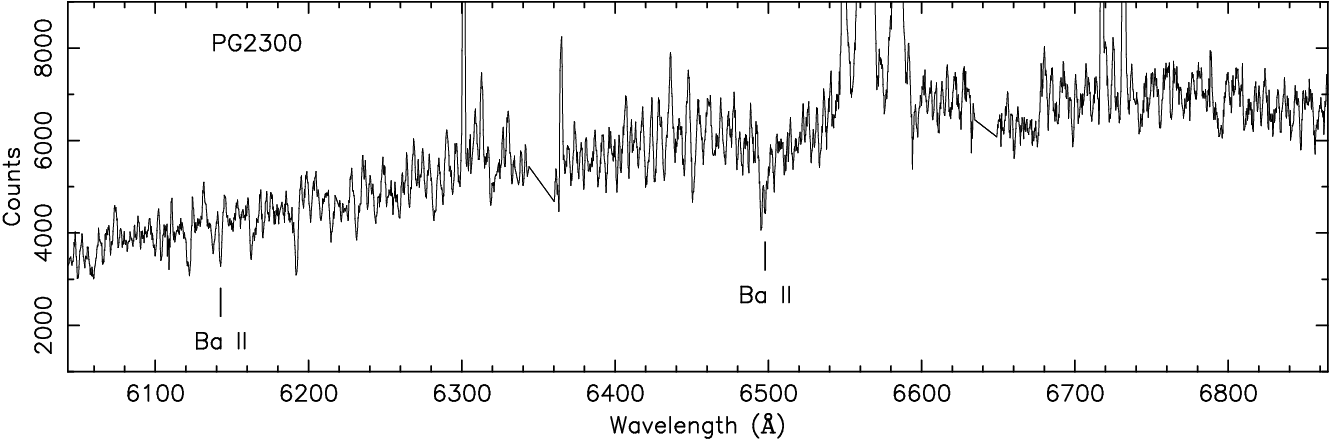}
   \end{center}
   \caption{SALT RSS spectra of the nucleus of Hen~2-39. The flux scale corresponds to $V=16.47$ mag (see Sect. \ref{sec:cspn}). No nebular emission was subtracted from the PG2300 spectra and no flux calibration was applied for the second PG2300 spectrum.}
   \label{fig:saltspec}
\end{figure*}

\begin{figure*}
   \begin{center}
      \includegraphics[scale=0.53,angle=270]{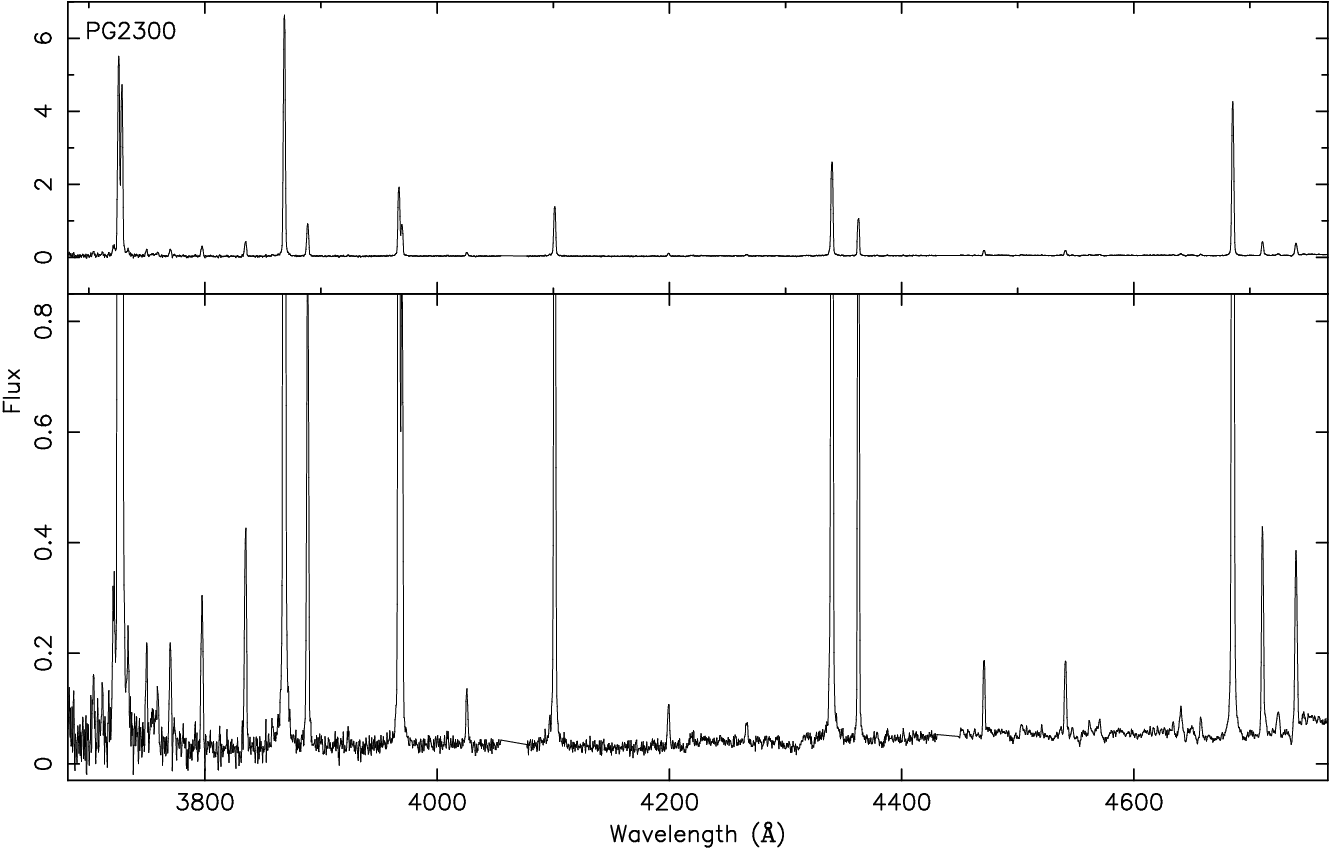}
      \includegraphics[scale=0.53,angle=270]{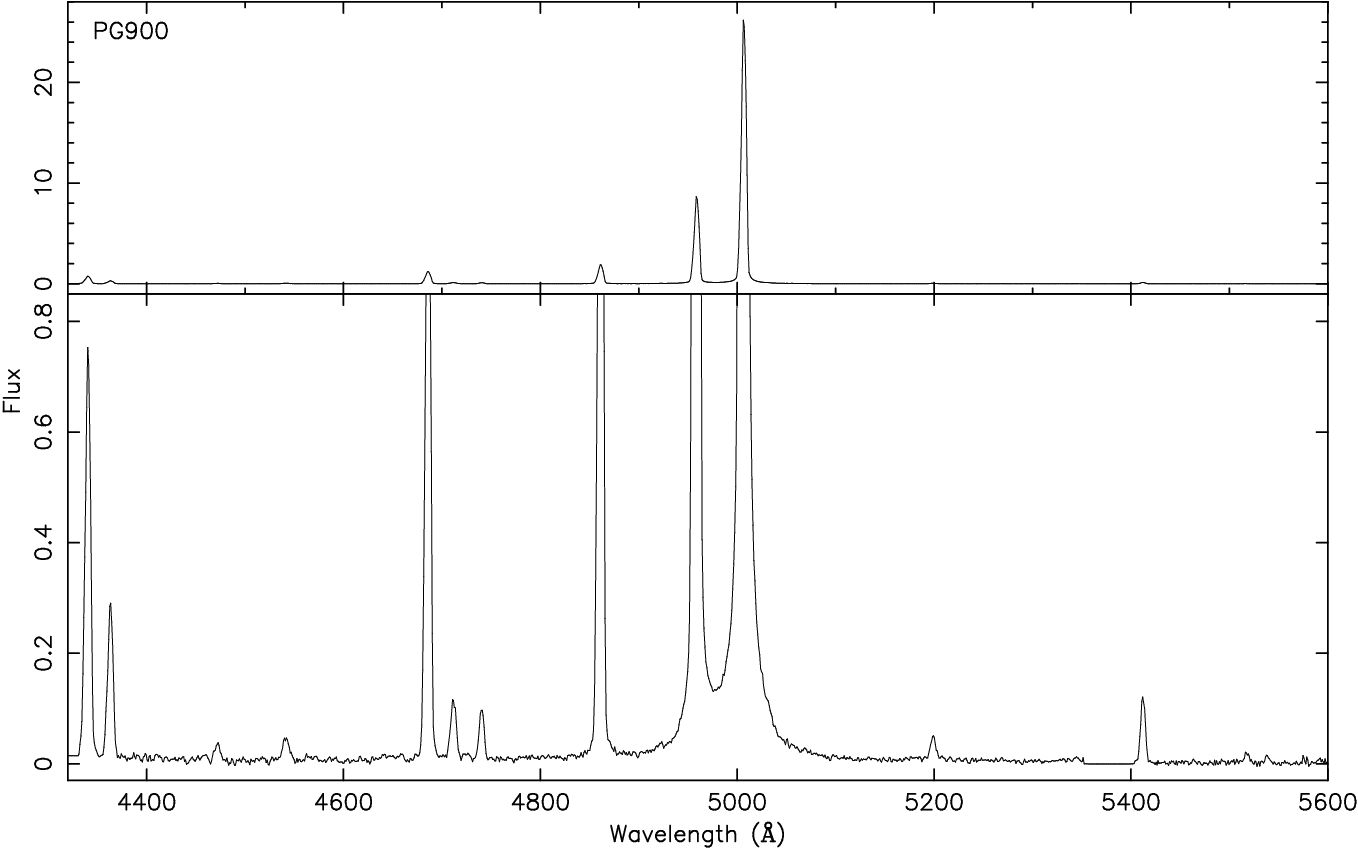}
      \includegraphics[scale=0.53,angle=270]{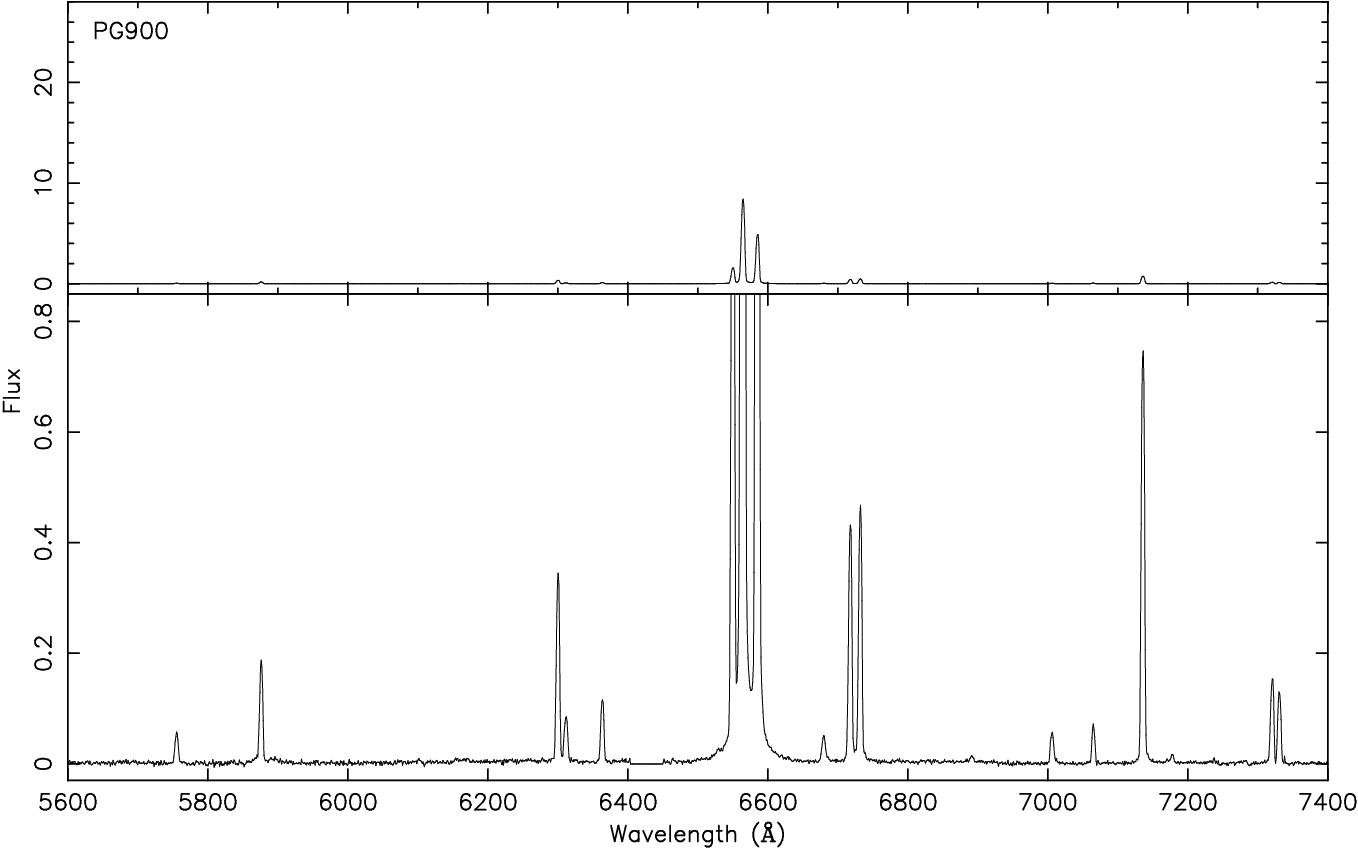}
   \end{center}
   \caption{SALT RSS nebular spectra.}
   \label{fig:neb}
\end{figure*}

A spectrum of the spectrophotometric standard star LTT 4364 (Hamuy et al. 1994) was used to flux calibrate the spectra in the usual fashion (excluding the last SALT spectrum). Due to the moving pupil design of SALT only a relative (not absolute) spectrophotometric solution can be obtained. Analysis of emission line intensities provides an independent check of the relative flux calibration (Sect. \ref{sec:neb}). 
   The resultant stellar and nebular spectra, discussed in the remainder of the paper, are depicted in Figures \ref{fig:saltspec} and \ref{fig:neb}, respectively.

\begin{table*}
   \centering
   \caption{Log of SALT RSS observations.}
   \label{tab:obs}
   \begin{tabular}{lllllll}
      \hline
        Date       & Grating & Exptime & Slit Width & $\lambda$ & $\Delta\lambda$ & Dispersion \\
        (DD/MM/YY) &         &  (s)    & ('')       & (\AA)     & (FWHM, \AA)     & (\AA\ pix$^{-1}$) \\
      \hline
       29/01/12 & PG2300 & 1500 & 1.25 & 3682--4767 & 1.8 & 0.35 \\
       16/02/12 & PG900 & 1500 & 1.25 & 4330--7404 &  5.0 & 0.97 \\
       27/05/12 & PG2300 & 2100 & 1.50 & 6042--6865 & 1.5 &  0.27 \\
      \hline
   \end{tabular}
\end{table*}

\subsection{Imaging}
\label{sec:img}

\begin{figure}
      \begin{center}
\includegraphics[scale=0.405]{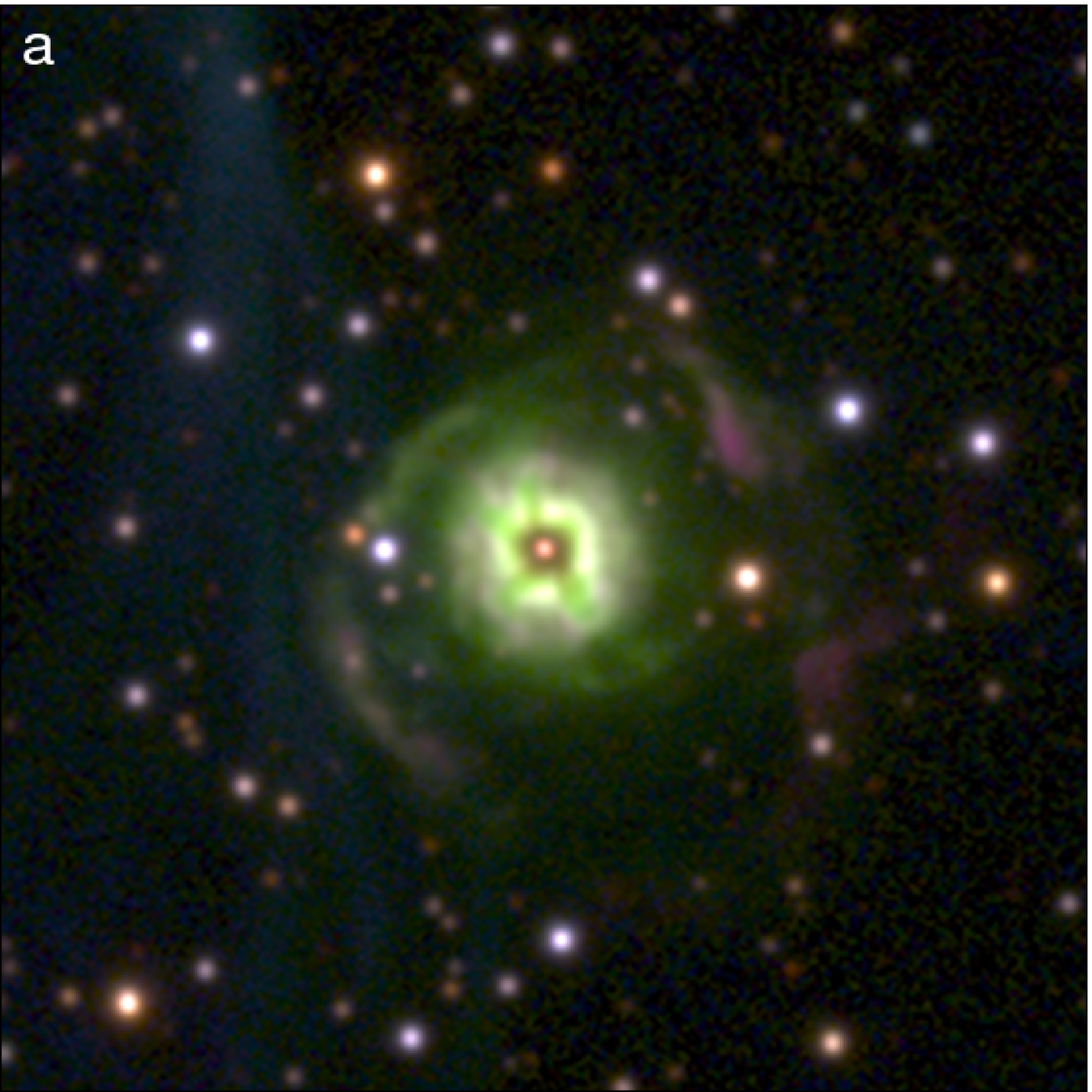}\\
\includegraphics[scale=0.2]{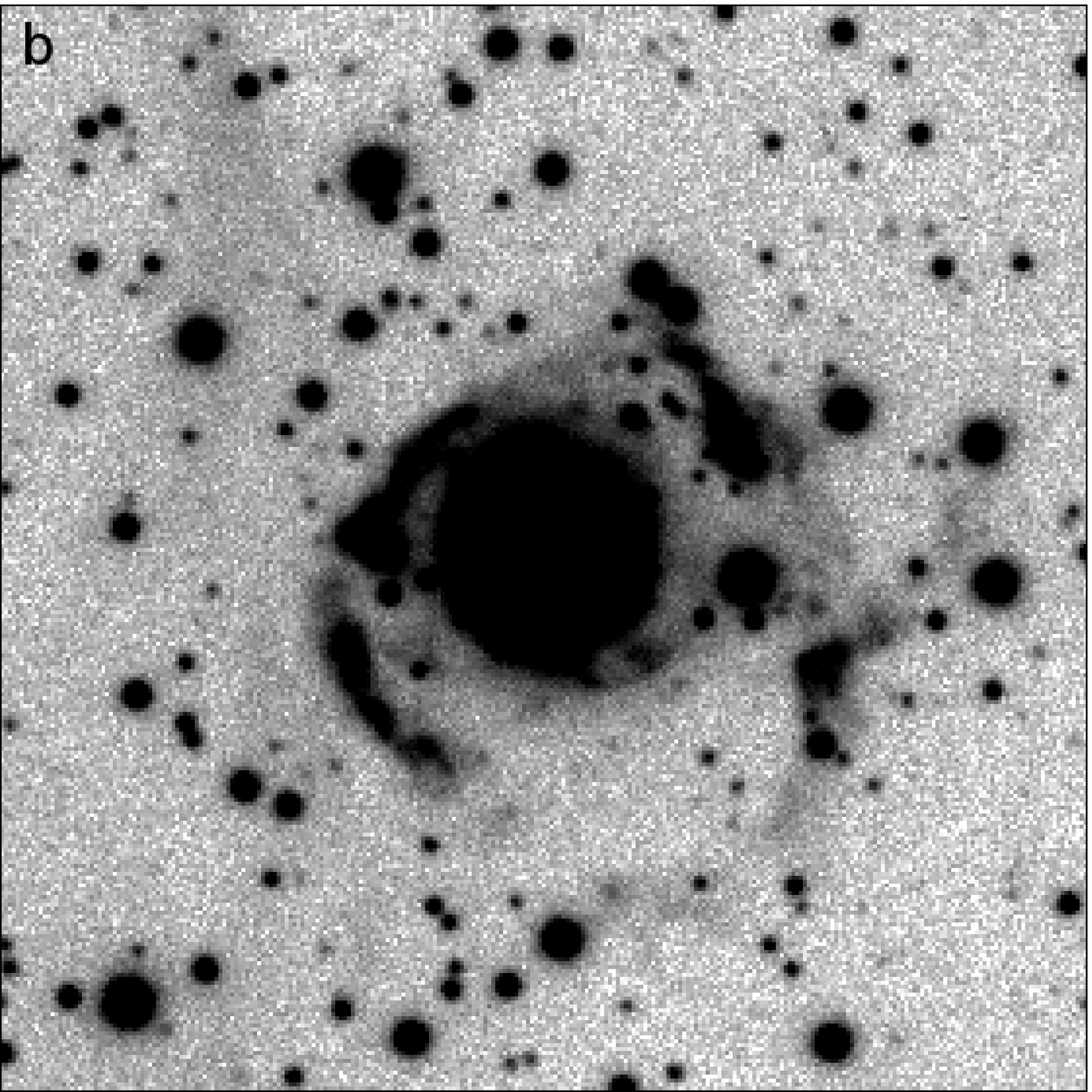}
\includegraphics[scale=0.2]{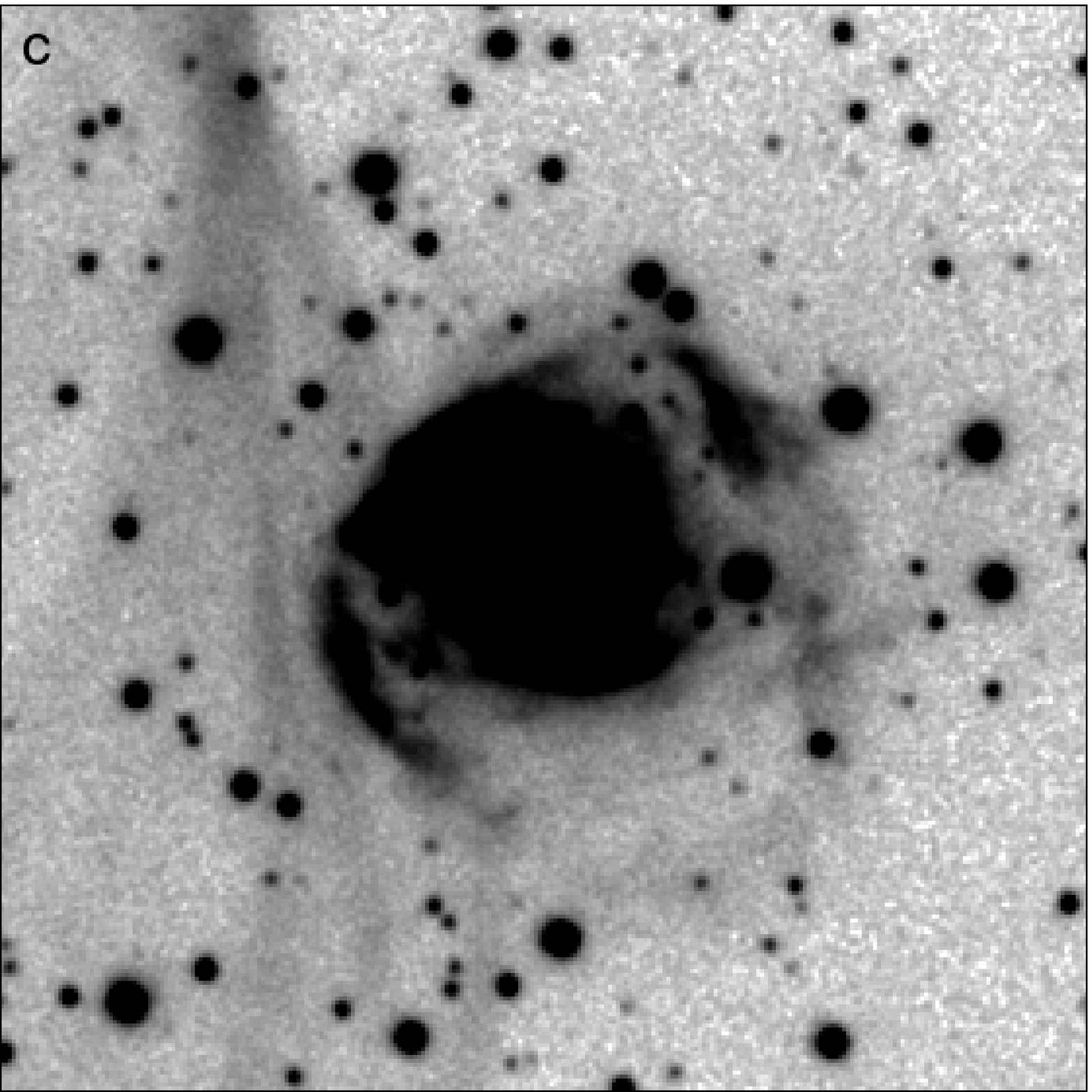}\\
\includegraphics[scale=0.2]{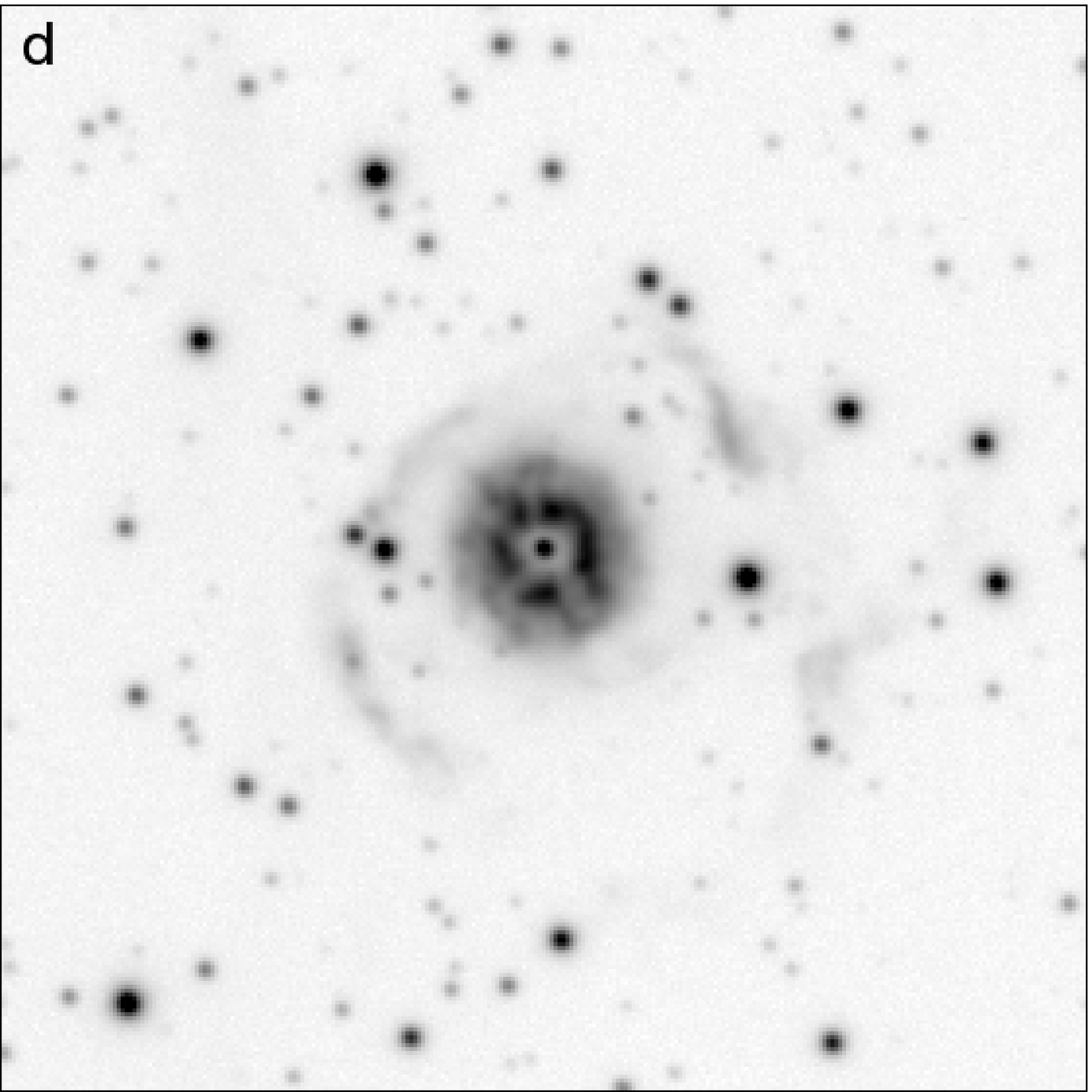}
\includegraphics[scale=0.2]{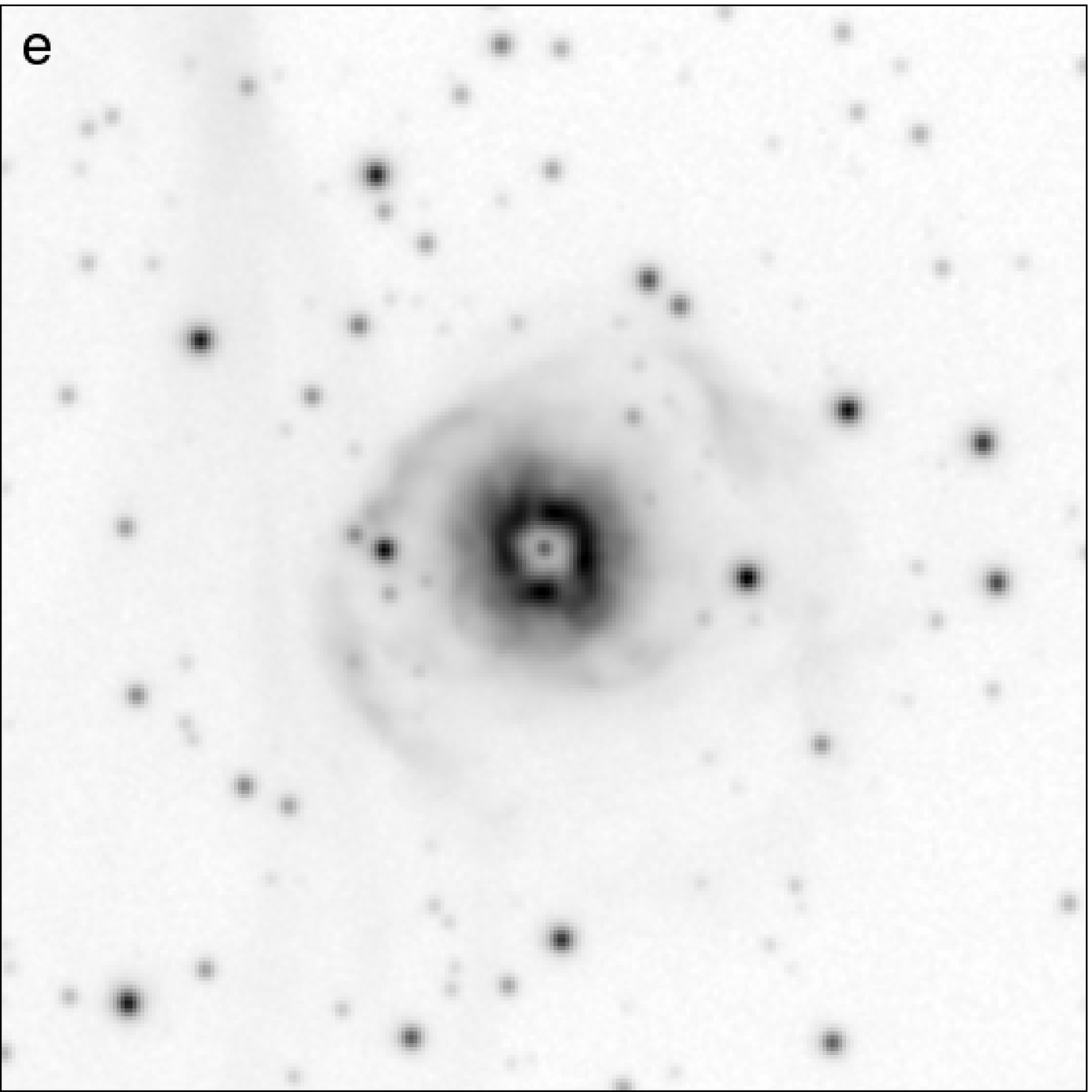}\\
\includegraphics[scale=0.2]{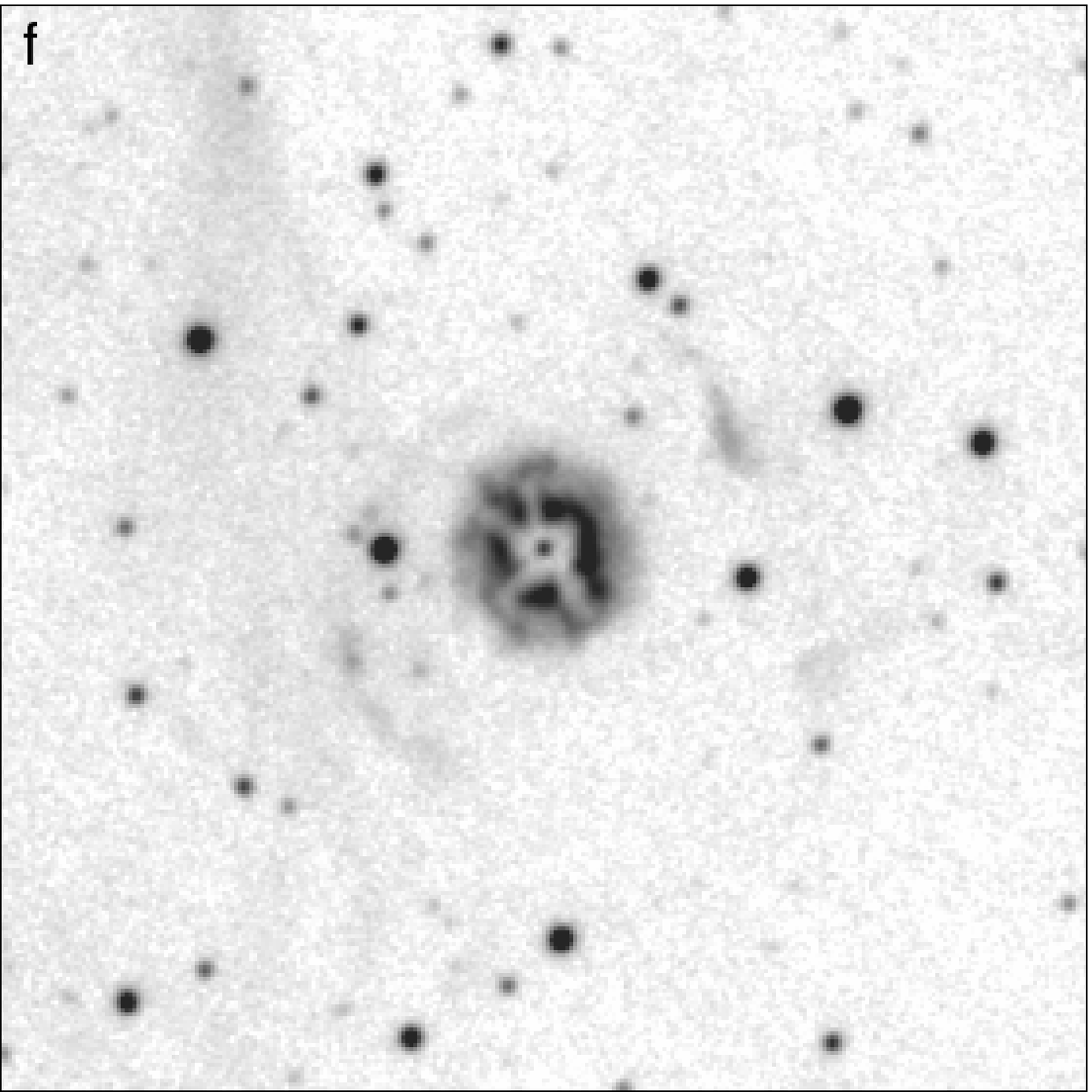}
\includegraphics[scale=0.2]{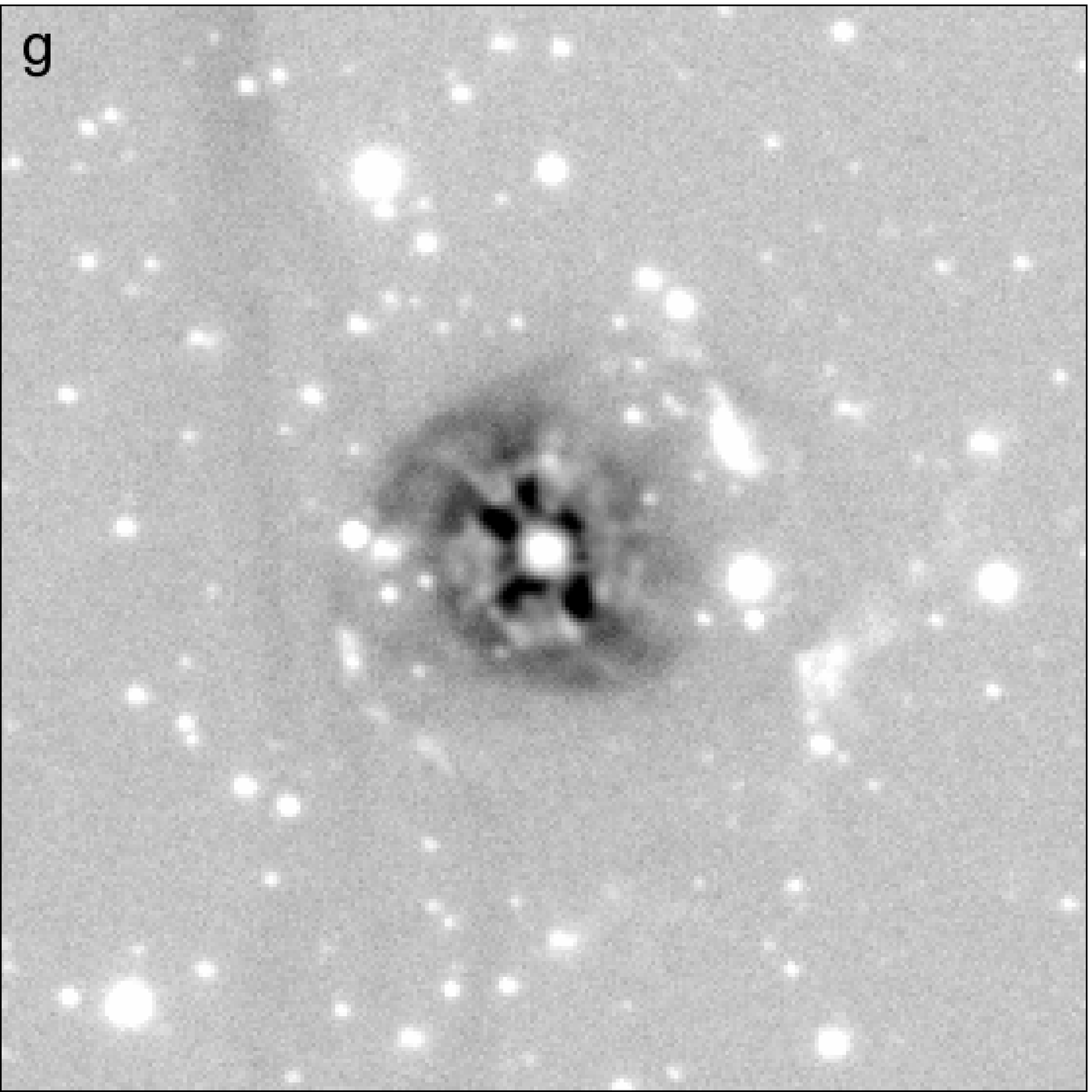}
      \end{center}
      \caption{VLT FORS2 images of Hen~2-39. (a) Colour-composite image made from H$\alpha$+[N~II] (red), [O~III] (green) and [O~II] (blue); (b) and (d) H$\alpha$+[N~II]; (c) and (e) [O~III]; (f) [O~II]; and (g) H$\alpha$+[N~II] divided by [O~III]. Image dimensions are $70\times70$ arcsec$^2$ with North up and East to left. Note the very red appearance of the central star in (a). The unrelated vertical filament just to the East of Hen~2-39 is ionized gas from the interstellar medium.}
      \label{fig:img}
   \end{figure}

Narrow-band images were taken with the focal reducer and low-dispersion spectrograph (FORS2; Appenzeller et al. 1998) on the Very Large Telescope (VLT) under the visitor mode programme 088.D-0750(A) on 12 February 2012. Narrow-band images of 180s duration each were taken with the H\_Alpha+83, OIII+50 and OII+44 filters. The central wavelengths and FWHM of each filter are 656.3/6.1 nm (includes [NII] $\lambda$6548, 6583), 500.1/5.7 nm and 377.6/6.5 nm, respectively. The data have a pixel scale of 0.252\arcsec\ pixel$^{-1}$ and were bias-subtracted and flat-fielded using the ESO FORS pipeline. Figure \ref{fig:img} presents a montage of the images that are discussed in Sect. \ref{sec:neb}.

   \subsection{Photometry}
   We monitored the central star of Hen~2-39 with the SAAO 1.9-m (24--28 March 2010) and 1.0-m telescopes (28 March to 3 April 2012 and 23--29 May 2012), and the ESO 3.6-m NTT (15--19 January 2013) under programme ID 090.D-0693(A). The SAAO CCD camera with the 1K$\times$1K STE4 CCD and a standard Cousins I-band filter ($\lambda$$\sim$700--900 nm) was used in direct imaging mode (binned $2\times2$) providing pixel scales of 0.28\arcsec\ pixel$^{-1}$ (1.9-m) and 0.62\arcsec\ pixel$^{-1}$ (1.0-m). At the NTT the EFOSC2 instrument (Buzzoni et al. 1984; Snodgrass et al. 2008) was used with the i\#705 (Gunn $i$) filter and an E2V CCD (\#40) with a pixel scale of 0.24\arcsec\ pixel$^{-1}$.  The SAAO observing strategy was to take whenever possible a set of contiguous exposures each time Hen~2-39 was observed, including often a second or third set of contiguous exposures taken during the same night, to maximise sensitivity to a range of variability timescales. Some weak nebular detection surrounding the central star is seen, particularly in the SAAO data that experienced poorer seeing. In order to minimise its influence, we used a fixed aperture of 4.0\arcsec\ radius when extracting the differential photometry of the central star, relative to two non-variable field stars, using \textsc{sextractor} (Bertin \& Arnouts 1996). This follows the strategy for minimising the seeing-dependent nebular influence on the photometry outlined in Jones (2011) and Miszalski et al. (2011) where the most effective aperture to use is roughly 1.5 times the poorest seeing of the observations (2.7\arcsec in this case). All the observations were reduced in the same manner and a few poorer quality frames were discarded. A total of 137 SAAO and 38 NTT measurements are included in the lightcurve with exposure times ranging from 20--60 s (NTT) to several minutes (SAAO).

\section{The central star}
\label{sec:cspn}
\subsection{Stellar properties}
\label{sec:basic}
The SALT spectra in Fig. \ref{fig:saltspec} reveal the central star to be a cool star dominated by molecular bands of CH, CN and C$_2$ (e.g. Davis 1987). Also prominent are Ba~II $\lambda$4554.03 and $\lambda$6496.90 absorption lines, the latter of which we use to determine an enhanced abundance relative to the Solar abundance (see below), distinguishing the cool star as a barium star. Using the Barnbaum atlas of carbon stars (Barnbaum, Stone \& Keenan 1996), based on the revised MK classification of red carbon stars (Keenan 1993), we classify the star as C-R3 C$_2$4, with an uncertainty of one class in the spectral type. We discuss later in Sect. \ref{sec:distance} the possible overlap between Hen~2-39 and early R-type carbon stars, an unusual subset of carbon stars. It closely resembles HD223392 in the Barnbaum atlas and WeBo~1 (Bond et al. 2003). Weak Ca~II K emission is detected, indicating some chromospheric activity may be present, but unfortunately the strong nebular [Ne~III] $\lambda$3967 and H~I $\lambda$3970 emission precluded any measurement of Ca~II H emission (as seen in WeBo~1, Bond et al. 2003). Similarly, we are unable to tell whether there is any chromospheric H$\alpha$ emission as in A~70 (Miszalski et al. 2012; Tyndall et al. 2013). At the bluest wavelengths in the first PG2300 spectrum, the stellar continuum is weak until $\sim$4200 \AA, as may be expected from the Bond \& Neff (1969) effect which describes a flux depression near 4000 \AA\ in barium stars. 

\begin{figure}
   \begin{center}
      \includegraphics[scale=0.38,angle=270]{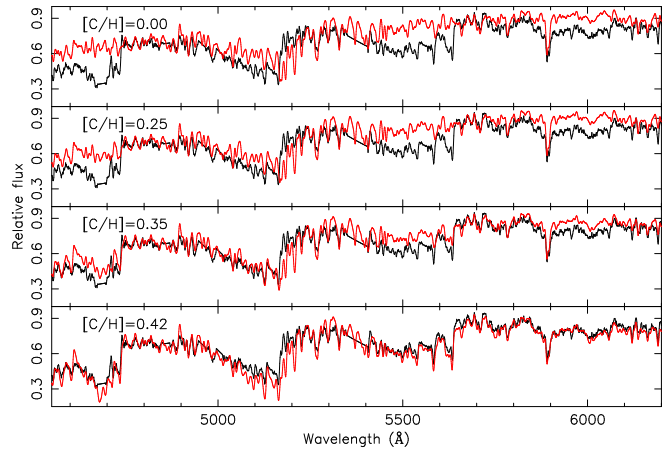}
   \end{center}
   \caption{The effect of increasing the model atmosphere carbon abundance compared to the observed SALT spectrum (black lines) for values of [C/H]=0.00, 0.25, 0.35 and 0.42 dex (red lines). The best fit is obtained for [C/H]=0.42 dex.}
   \label{fig:carbon}
\end{figure}

\begin{figure}
   \begin{center}
      \includegraphics[scale=0.38,angle=270]{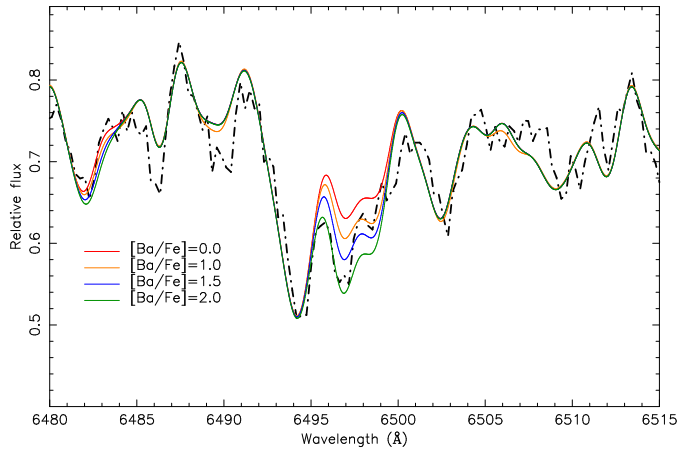}
   \end{center}
   \caption{The effect of increasing the model atmosphere barium abundance (coloured lines) compared to the observed SALT spectrum (black dash-dotted line) for values of [Ba/Fe]=0.0, 1.0, 1.5 and 2.0 dex. The best fit value to Ba~II $\lambda$6496.90 is [Ba/Fe]=$1.50\pm0.25$ dex.}
   \label{fig:barium}
\end{figure}

\begin{figure}
   \begin{center}
      \includegraphics[scale=0.38,angle=270]{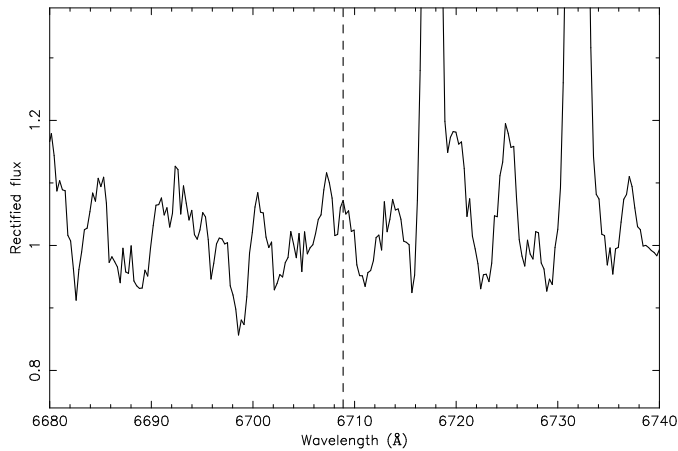}
   \end{center}
   \caption{The region around Li I 6707.83 in the second PG2300 spectrum. The dashed vertical line indicates the expected position of the line at the radial velocity of Hen~2-39.}
   \label{fig:lithium}
\end{figure}

We analysed the low-resolution spectrum with the stellar spectral synthesis code of R. Gray, \textsc{spectrum}\footnote{http://www1.appstate.edu/dept/physics/spectrum/spectrum.html} version 2.76, with models from Castelli \& Kurucz (2006). The best fit parameters were determined to be $T_\mathrm{eff}=4250\pm150$ K and log $g=2.0\pm0.5$, nominally corresponding to a K3III spectral type. Siegel et al. (2012) determined similar values of $T_\mathrm{eff}=4750$ K and log $g=2.0$ for WeBo~1. A significant carbon enhancement of [C/H]=0.42$\pm$0.02 dex is required to fit the observed spectrum of Hen~2-39 (Fig. \ref{fig:carbon}). The iron abundance was found to be close to solar when compared to the observed spectra, while oxygen is also assumed to be solar since it is not well constrained from CO bands. To estimate the barium abundance we used the Ba~II $\lambda$6496.90 line, since the low S/N of the Ba~II 4554.03 and  6141.71 lines did not allow for a reliable analysis. Figure \ref{fig:barium} displays the region around the Ba~II $\lambda$6496.90 line and several model atmospheres with varying [Ba/Fe] abundance from which we estimate [Ba/Fe]=$1.50\pm0.25$ dex. In this spectrum the heliocentric RV of the two Ba~II lines ($42.2\pm1.0$ km s$^{-1}$) is consistent with the nebular emission lines (36.7$\pm$4.5 km s$^{-1}$) measured by \textsc{emsao} (Kurtz \& Mink 1998), proving the physical association between the carbon star and the nebula (allowing for some small difference due to orbital motion). Furthermore, we find no evidence for Li I $\lambda$6707.83 in the second PG2300 spectrum (Fig. \ref{fig:lithium}), consistent with many Galactic carbon stars being lithium poor (Abia et al. 1993).

Table \ref{tab:stellar} summarises the stellar properties and magnitudes of the observed nucleus of Hen~2-39. Tylenda et al. (1991) measured $B$ and $V$ magnitudes, while NIR magnitudes were measured by DENIS (Epchtein et al. 1999) and 2MASS (Skrutskie et al. 2006). We can make use of the DENIS $I$-band magnitude to check the Tylenda et al. (1991) measurements. The design of SALT precludes measurement of magnitudes from the SALT spectra (see Sect. \ref{sec:obs}), and our wavelength coverage was too small to calculate a relative $V-I$ colour. Instead, we used the \textsc{synphot} package developed by STScI\footnote{http://www.stsci.edu/institute/software\_hardware/stsdas/synphot} to measure an intrinsic $V-I$ colour of $(V-I)_0=1.22$ mag from our model atmosphere spectrum. This is in good agreement with $(V-I)_0=1.26$ for a normal giant of the same $T_\mathrm{eff}$ (Bessell 1979). Starting with $E(B-V)=0.37$ derived from our nebular analysis (see Sect. \ref{sec:neb}) and the observed DENIS $I$-band magnitude, we calculated $A_V=3.1\ E(B-V)=1.13$ mag and $A_I=0.5\ A_V=0.57$ mag (Cardelli, Clayton \& Mathis 1989), allowing for the intrinsic $I$ and $V$ magnitudes to be calculated. This resulted in an observed $V=16.47$ mag, in very good agreement with Tylenda et al. (1991). Following a similar process we derive $U=20.69$ mag, $B=18.38$ mag, somewhat fainter than $B=17.9$ mag from Tylenda et al. (1991), and $R=15.52$ mag. The difference between the $B$ magnitudes might be construed as a contribution from the pre-WD companion, however this interpretation is not substantiated by the high error of 0.25--0.50 mag in the Tylenda et al. (1991) $B$ magnitude (see also Sect. \ref{sec:wd}). High spatial resolution $U$ and $B$ observations may in the near future be able to detect any excess if the pre-WD contributes a small fraction to the total light of the central star.

\begin{table}
   \centering
   \caption{Properties of the observed polluted giant secondary of Hen~2-39. Model dependent quantities rely upon measurements made from the model atmosphere described in the text.} 
   \label{tab:stellar}
   \begin{tabular}{llll}
      \hline
      $T_\mathrm{eff}$ & $4250\pm150$ & K & this work\\
      log $g$ & $2.0\pm0.5$ & cm s$^{-2}$& this work\\
      Type & C-R3 C$_2$4 & & this work\\
      {}[Ba/Fe] & $1.50\pm0.25$ & dex & this work\\
      {}[C/H] & $0.42\pm0.02$ & dex & this work\\
      $B$ & $17.9\pm0.5$ & mag & Tylenda et al. (1991)\\
      $V$ & $16.5\pm0.5$ & mag & Tylenda et al. (1991)\\
      $U$ & $20.69$  & mag & this work (model dep.)\\
      $B$ & $18.38$ & mag & this work (model dep.)\\
      $V$ & $16.47$ & mag & this work (model dep.)\\
      $R$ & $15.52$  & mag & this work (model dep.)\\
      $(U-B)_0$ & 2.05 & mag & this work (model dep.)\\
      $(B-V)_0$ & 1.53 & mag & this work (model dep.)\\
      $(V-R)_0$ & 0.67 & mag & this work (model dep.)\\
      $(V-I)_0$ & 1.22 & mag & this work (model dep.)\\
      $(V-Ks)_0$ & 3.12 & mag & this work\\
      $I$ & $14.68\pm0.03$ & mag & Epchtein et al. (1999)\\
      $J$ & $13.37\pm0.09$ & mag & Epchtein et al. (1999)\\
      $K$ & $12.40\pm0.12$ & mag & Epchtein et al. (1999)\\
      $J$ & $13.47\pm0.03$ & mag & Skrutskie et al. (2006)\\
      $H$ & $12.61\pm0.03$ & mag & Skrutskie et al. (2006)\\
      $K_s$ & $12.34\pm0.03$ & mag & Skrutskie et al. (2006)\\
      $J-H$ & $0.86$ & mag & Skrutskie et al. (2006)\\
      $H-K_s$ & $0.28$ & mag & Skrutskie et al. (2006)\\
      \hline
   \end{tabular}
\end{table}

\subsection{Photometric Variability}
\begin{figure}
   \begin{center}
      \includegraphics[scale=0.35,angle=270]{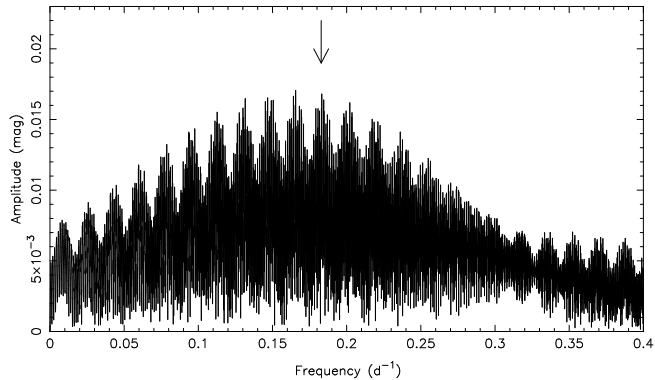}
   \end{center}
   \caption{Fourier periodogram of the combined lightcurve calculated by the \textsc{period04} package. The arrow indicates the peak at $f=0.182$ d$^{-1}$ corresponding to $5.46$ d, but other periods may also be plausible (see text).}
   \label{fig:fourier}
\end{figure}

The most likely origin for any periodic variation in a red giant star are short-lived spots at the surface of the star that move with the rotation period. We have therefore, as a first approach, treated the two datasets (SAAO and NTT) separately and searched for any periodicities using version 1.2 of the \textsc{period04}\footnote{http://www.univie.ac.at/tops/Period04} fourier analysis package (Lenz \& Breger 2004). The SAAO data have a zeropoint of $-0.316$ mag and an initial root-mean-square (RMS) dispersion of 0.0142 mag. The periodogram provides several periods whose associated sinusoidal amplitude is between 0.013--0.016 mag, and which decrease the RMS dispersion to 0.0096 mag. The S/N of these peaks are only about 3 and are therefore formally not significant. The frequencies that provide the best folded lightcurves are: 0.147856, 0.094985 and 0.183138 d$^{-1}$, as well as their 1 d aliases. Figure \ref{fig:fourier} shows the very broad peaks of the periodogram, meaning any exact value of the frequency is difficult to obtain.

\begin{figure}
   \begin{center}
      \includegraphics[scale=0.47]{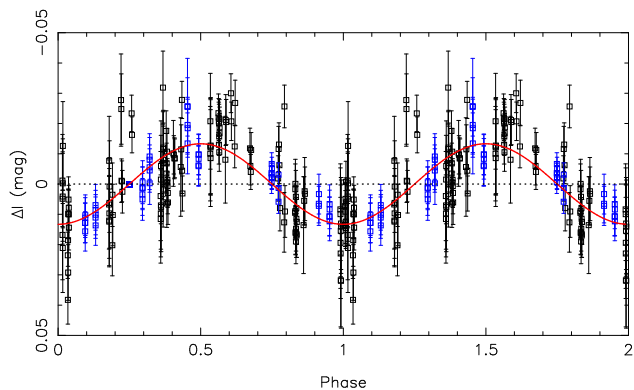} 
   \end{center}
   \caption{Combined lightcurve phased with the 5.46 d period. SAAO and NTT observations are represented as black and blue squares, respectively, and the red line is a sinusoid of amplitude 0.0133 mag.}
   \label{fig:phased}
\end{figure}

It is reassuring that similar frequencies are obtained when looking at the periodogram from the NTT data alone. This data set has a zero point of 0.057 mag (different to the SAAO zeropoint because of different filters) and an initial RMS dispersion of 0.019 mag. When looking in particular around the above frequencies, we find that the 0.147876 and 0.182891 d$^{-1}$ frequencies lead to the smallest residuals, around 0.0086 and 0.0089 mag, respectively. A decrease of more than a factor 2 in the residual is a significant result. Phase folding the NTT data with the 0.094985 frequency does not give any valid result, and we can therefore reject this frequency. We finally combined each data set in which we rescaled the SAAO and NTT data with the amplitude and phase found for the solutions of the two remaining frequencies before recalculating the periodogram. The result is that the 0.182891 d$^{-1}$ frequency (corresponding to a period of 5.46 d) provides the best match and the most agreeable phase folded lightcurve (Fig. \ref{fig:phased}). We are tentatively led to conclude that the rotation period of the star is around 5.5 days, even though we cannot formally reject any periods between 4.5 and 11 days. 

The periodic lightcurve indicates that the apparent nucleus of Hen~2-39 is a fast rotating giant, consistent with the periods found for similar stars. These include 5.9 d for LoTr~5 (Jasniewicz et al. 1996; Strassmeier, Hubl \& Rice 1997) and 4.7 d for WeBo~1 (Bond et al. 2003). A rotation period of 6.4 d was determined for LoTr~1 (Jones et al. 2011; Tyndall et al. 2013), however Tyndall et al. (2013) found LoTr~1 not to be a barium star due to an insufficient overabundance of barium. These stars are suspected to be spun-up during the wind accretion phase (e.g. Theuns, Boffin \& Jorissen 1996; Jeffries \& Stevens 1996; Montez et al. 2010).

\section{The nebula}
\label{sec:neb}
\subsection{Apparent morphology}
Figure \ref{fig:img} reveals Hen~2-39 to be a fragmented inner ring surrounded by several outer filaments. A detailed spatiokinematic analysis will be needed to determine the intrinsic morphology of Hen~2-39, however its ring-like appearance does seem to fit the trend of ring nebulae surrounding barium CSPNe (Bond et al. 2003; Miszalski et al. 2012; Tyndall et al. 2013). Recent spatiokinematic modelling by Tyndall et al. (2013) has shown both WeBo~1 and A~70 to have a bipolar morphology viewed pole-on (e.g. Jones et al. 2012). A bipolar shape was also determined for LoTr~5 (Graham et al. 2004). The diameter of the inner nebula was measured from a contour of 10 per cent of the maximum intensity of the inner nebula to be 13.1\arcsec\ (H$\alpha$ and [O~II]). The dimensions are asymmetric in [O~III] where we measured $13.1\times14.6$\arcsec. The SW filament appears to be pinched, perhaps suggesting the presence of a bipolar envelope to Hen~2-39. Another alternative may be that the filaments represent a typical recombination halo (Corradi et al. 2003). The brightest filaments of the halo lie 16.2\arcsec\ from the central star and the outermost edges of the SW filament 24.9\arcsec.

   \begin{table}
      \centering
      \caption{Basic nebula properties of Hen~2-39.}
      \label{tab:basic}
      \begin{tabular}{lrl}
         \hline
         PN G & 283.8$-$04.2 &\\
         RA & 10$^\mathrm{h}$03$^\mathrm{m}$49\fs2 & \\
         Dec. & $-$60$^\circ$43$'$50\arcsec &\\
         Nebula diameter & 13.1\arcsec & this work\\
         Maximum halo extent & 24.9\arcsec & this work\\
         $\log F(H\beta)$ & $-12.4\pm0.2$ & Acker et al. (1992)\\ 
$v_\mathrm{hrv}$ & $36.7\pm4.5$ km s$^{-1}$ & this work\\ 
         $d$ & 5.7 kpc & this work\\
         $z$ & $-420$ pc & this work\\
         \hline
      \end{tabular}
   \end{table}

   \subsection{Nebular plasma parameters and abundances}
   \label{sec:chem}

Emission line fluxes were measured from the flux calibrated SALT spectra covering $\lambda$3682--4767\AA\ (PG2300) and 4330--7404\AA\ (PG900) using the \textsc{iraf} task \textsc{splot}. The brightest line in the overlap region, He~II $\lambda$4686, was used to match the two sets of line fluxes together. This resulted in an agreement of better than 0.5 per cent of H$\beta$ between line fluxes in the overlap region. Table \ref{tab:linelist} contains the composite nebular spectrum of Hen~2-39. Since the [O~III] $\lambda$5007 line was slightly saturated, we exclude it from our subsequent analysis. The nebular CIII and NIII emission lines could not be reliably measured because of their faintness and small spatial extent that means they were more affected by the non-uniform stellar continuum. 

The line fluxes in Tab. \ref{tab:linelist} were analysed with the plasma diagnostics program \textsc{hoppla} (Acker et al. 1991; K\"oppen et al. 1991; Girard et al. 2007) in a similar fashion to Miszalski et al. (2012). It  first determines the extinction constant, using the reddening law of Howarth (1983), as well as the electron 
temperature and density in a consistent way. 
The S/N of the temperature sensitive [O~III] $\lambda$4363 and [N~II] $\lambda$5755 emission lines were 65 and 19, respectively. 
All seven Balmer lines fit within 3 per cent of their intensity from case B recombination theory.
The He~I line intensities are matched better than 10 per cent by case B 
recombination emissivities corrected for collisional excitation 
(Porter, Ferland \& MacAdam, 2007), except the $\lambda$4471 line, which is 
too weak by 16 per cent.

  \begin{table}\centering
     \caption[]{The nebular spectrum of Hen~2-39 combined from the 
      fluxes of the PG2300 and PG900 SALT spectra. Fluxes and dereddened emission line intensities 
      are scaled to H$\beta=100$.} 
      \label{tab:linelist}
     \begin{tabular}{lrr}
     \noalign{\smallskip} \hline  \noalign{\smallskip}
           Line  & Flux   & Intensity  \\
     \noalign{\smallskip} \hline
     \noalign{\smallskip}
[O II] 3726         &  81.7   &  111.4 \\
{}[O II] 3729       &  67.5   &   92.0 \\
{}H I 3750          &   2.2   &    3.0 \\
{}H I 3771          &   2.9   &    3.9 \\
{}H I 3798$^a$      &   4.4   &   5.9  \\
{}H I 3835$^a$      &   6.4   &   8.5  \\
{}[Ne III] 3869     &  99.8   &  131.9 \\
{}He I 3889$^a$     &  13.8   &   18.1  \\
{}[Ne III]+H I 3967 &  29.2   &   37.7 \\
{}H I 3970          &  12.1   &   15.6 \\
{}He I 4026$^a$     &   1.5   &   1.9  \\
{}H I 4102          &  21.3   &   26.5 \\
{}He II 4200        &   1.3   &    1.6 \\
{}C II 4267         &   0.7   &    0.8 \\
{}H I 4340          &  41.2   &   48.0 \\
{}[O III] 4363      &  15.3  &   17.8 \\ 
{}He I 4471         &   1.8   &    2.0 \\
{}He II 4541        &   2.4   &    2.6 \\
{}He II 4686        &  65.9   &   69.4 \\
{}[Ar IV]+He I 4711 &   6.3   &    6.6 \\
{}[Ne IV] 4724      &   1.0  &    1.0 \\ 
{}[Ar IV] 4740      &   4.7   &    4.9 \\
{}H I 4861          & 100.0   &  100.0 \\
{}[O III] 4959      & 452.9   &  440.0 \\
{}[O III] 5007$^a$  &1369.7   &    1311.8 \\
{}[N I] 5200        &   2.4   &    2.2 \\
{}He II 5412        &   5.9   &    5.0 \\
{}[Cl III] 5518     &   0.9   &    0.8 \\
{}[Cl III] 5538     &   0.6   &    0.5 \\
{}[N II] 5755       &   2.6   &    2.1 \\
{}He I 5876         &   8.6   &    6.6 \\
{}[O I] 6300        &  15.9   &   11.3 \\
{}[S III] 6312      &   4.1   &    2.9 \\
{}[O I] 6363        &   5.1   &    3.6 \\
{}[N II] 6548       &  77.2   &   52.6 \\
{}H I 6563          & 416.0   &  282.4 \\
{}[N II] 6583       & 234.5   &  159.7 \\
{}He I 6678         &   2.4   &    1.6 \\
{}[S II] 6717       &  20.8   &   13.8 \\
{}[S II] 6731       &  22.0   &   14.5 \\
{}He II 6891        &   0.5   &    0.3 \\
{}[Ar V] 7005       &   2.8   &    1.8 \\
{}He I 7065         &   3.0   &    1.9 \\
{}[Ar III] 7135     &  36.1   &   22.4 \\
{}He II  7178       &   0.7   &    0.4 \\
{}[O II] 7320       &   7.3   &    4.4 \\
{}[O II] 7330       &   6.0   &    3.6 \\
    \noalign{\smallskip} \hline
  \end{tabular}
  \begin{flushleft}
     $^a$Line not used in \textsc{hoppla} analysis.
  \end{flushleft}
\end{table}

Several diagnostic line ratios sensitive to electron densities are available. 
The blue [O~II] lines are resolved and yield 820~cm$^{-3}$
and the red [S~II] lines 755~cm$^{-3}$. We adopt a value of 787~cm$^{-3}$
as the average of the intensity-weighted logarithmic individual densities.
The density may also be measured using the ratio [Ar~IV] 4711/4740, that 
we calculated to be 1.29 (after subtracting the He~I $\lambda$4713.2 intensity
of 0.3 per cent of H$\beta$ from the intensity of $\lambda$4711), 
corresponding to a density of 490~cm$^{-3}$. As this is in the relatively 
flat part near the low density limit, it would also be in acceptable agreement
with our adopted value. The [Cl~III] line ratio is at its low-density limit, but these lines are very faint at $<$1 per cent of H$\beta$. Thus, all four density indicators give consistent values.

Using the electron density and the temperatures for low, middle, and high ionic 
species (given in Tab. \ref{tab:plasma}), ionic abundances were derived from the lines 
listed in Tab. \ref{tab:ionic} and weighted with the emissivity of individual lines. 
Elemental abundances are then obtained by summing over the ionic abundances 
and correcting for unobserved stage of ionization (see Miszalski et al. 2012 
for details). The resulting abundances are listed in Tab. \ref{tab:plasma}, together with the abundances for Abell~70 (Miszalski et al. 2012), average abundances for Type I and non-Type I PNe (Kingsburgh \& Barlow 1994) and Solar abundances (Asplund et al. 2009) for comparison.

  \begin{table}\centering
     \caption[]{Ionic abundances (relative to H$^+$) and the emission 
        lines from which they were derived.} 
        \label{tab:ionic}
     \begin{tabular}{lll}
     \noalign{\smallskip} \hline  \noalign{\smallskip}
      Ion    & Abundance & Emission lines used   \\
     \noalign{\smallskip} \hline
     \noalign{\smallskip}
      He$^+$    & $4.83\times10^{-2}$  & 5876, 4471, 6678, 7065  \\
      He$^{++}$ & $5.82\times10^{-2}$  & 4686, 5411, 4541, 4200, 7178 \\
      C$^{+}$   & $1.67\times10^{-3}$  & 4267 \\
      N         & $1.15\times10^{-6}$  & 5200 \\
      N$^+$     & $2.61\times10^{-5}$  & 6583, 6547, 5755 \\
      O         & $1.78\times10^{-5}$  & 6300, 6363 \\
      O$^+$     & $7.52\times10^{-5}$  & 3726, 3729, 7320, 7330 \\
      O$^{++}$  & $2.14\times10^{-4}$  & 4959, 4363 \\
      Ne$^{++}$ & $5.15\times10^{-5}$  & 3869 \\
      Ne$^{3+}$ & $2.20\times10^{-4}$  & 4724 \\
      S$^+$     & $6.45\times10^{-7}$  & 6717, 6731 \\
      S$^{++}$  & $5.93\times10^{-6}$  & 6312 \\
      Cl$^{++}$ & $4.00\times10^{-8}$  & 5517, 5537 \\
      Ar$^{++}$ & $1.20\times10^{-6}$  & 7135 \\
      Ar$^{3+}$ & $7.19\times10^{-7}$  & 4740 \\
      Ar$^{4+}$ & $1.78\times10^{-7}$  & 7005 \\
    \noalign{\smallskip} \hline
  \end{tabular}
\end{table}

  \begin{table}\centering
     \caption[]{Plasma diagnostics and the chemical composition of Hen 2-39. The abundances are 
     given in the usual notation $12+\log\left[n(\mathrm{ X})/n(\mathrm{ H})\right]$.
        } 
        \label{tab:plasma}
     \begin{tabular}{lrrrrr}
     \noalign{\smallskip} \hline \noalign{\smallskip}
                         & Hen~2-39 & non- &A~70 & Type I &  Sun \\
                         &          &  Type I    &     &        &      \\
     \noalign{\smallskip} \hline
     \noalign{\smallskip}
      $c$                         &  0.53 & --- & 0.07   & --- & --- \\
      $n_e$ [cm$^{-3}$]           &   787 & --- & 100    & --- & --- \\
      $T_e(\mathrm{O}^+)$ [K]     &  9740 & --- & 12400  & --- & ---\\   
      $T_e(\mathrm{O^{++}})$ [K]  & 13000 & --- & 13200  & --- & --- \\ 
      $T_e(\mathrm{He^{++}})$ [K] & 12660 & --- &12800   & --- & ---\\  
     \noalign{\smallskip} \hline
     \noalign{\smallskip}
      He    & 11.03  & 11.05 & 11.28 &  11.11& 10.93 \\
      C     &  9.22  & 8.81  & ---   &  8.43 & 8.43  \\
      N     &  8.35  & 8.14  & 8.68  &  8.72 & 7.83  \\
      O     &  8.80  & 8.69  & 8.43  &  8.65 & 8.69  \\
      Ne    &  8.19  & 8.10  & (7.68) &  8.09& 7.93  \\
      S     &  7.00  & 6.91  & 6.82  &  6.91 & 7.12  \\
      Cl    &  5.29  & ---   & 4.61  &  ---  & 5.50  \\
      Ar    &  6.37  & 6.38  & 6.11  &  6.42 & 6.40  \\
     \noalign{\smallskip} \hline
     \noalign{\smallskip}
      log(C/O)  & $+0.42 $ & $0.12$  & ---     & $-0.12$  & $-0.26$ \\
      log(N/O)  & $-0.45 $ & $-0.55$ & $+$0.25 & $+0.07$ & $-0.86$ \\
      log(Ne/O) & $-0.61 $ & $-0.59$ & ---     & $-0.56$ & $-0.76$ \\
      log(S/O)  & $-1.80 $ & $-1.78$ & $-$1.61 & $-1.74$ & $-1.57$ \\
      log(Cl/O) & $-3.51 $ & ---     & $-$3.82 & ---      & $-3.19$ \\
      log(Ar/O) & $-2.43 $ & $-2.31$ & $-$2.32 & $-2.23$ & $-2.29$ \\
    \noalign{\smallskip} \hline
  \end{tabular}
\end{table}

The derived abundances of Hen~2-39 reflect a composition close to Solar that is consistent with other non-Type I PNe. While the oxygen and neon abundances are slightly above Solar by 0.11 and 0.26 dex, respectively, their values may be affected by other processes (Karakas et al. 2009). The argon abundance is only marginally sub-Solar by 0.03 dex, and since argon is not expected to change during AGB nucleosynthesis (Karakas et al. 2009), it is a better tracer of the metallicity. We therefore conclude that the metallicity is Solar in Hen~2-39, in agreement with our model atmosphere analysis (Sect. \ref{sec:basic}). The N/O ratio is slightly higher than in the Sun at $-0.45$ dex and close to the average non-Type I PN. It is far too low for Hen~2-39 to be considered a Type I PN whose composition is defined by larger nitrogen enhancements (N/O $\ga$ $-$0.3, Peimbert \& Torres-Peimbert 1983; N/O $\ga$ $-$0.1, Kingsburgh \& Barlow 1994). This suggests the AGB star in Hen~2-39 had an initial mass $\la$2.5 $M_\odot$ (Kingsburgh \& Barlow 1994), but according to AGB model yields of N/O it could be as high as $\sim$4.5 $M_\odot$ (Karakas 2010) before N/O exceeds $-0.45$ as in Hen~2-39.

The $\alpha$-element ratios Ne/O, S/O, and Ar/O ratio differ slightly from the Solar values, but are very close to those of the average PN, and they agree very 
well with the average values of Ne/O $= -0.67$, S/O $= -1.55$, and Ar/O $= -2.25$
found in HII regions of spiral and irregular galaxies (Henry \& Worthey 1999).
Since these elements are not expected to be affected by the nucleosynthesis in the 
progenitor star, their ratios reflect the enrichment of the interstellar medium as determined by 
the nucleosynthesis in massive stars. Considering the extremely weak lines and the 
uncertainties of the ionization correction, chlorine is also considered Solar. 

The carbon abundance derived from the recombination line is 0.79 dex higher than in the Sun. This reflects the well-known effect that recombination lines yield
systematically higher abundances than collisionally excited lines (e.g. Liu et al. 2006). If one wanted to match the Solar C/O ratio, and ignoring any ionization 
factors, one would need an enhancement factor of 5, which is not unreasonably high. UV spectroscopy is required to better constrain the nebular carbon abundance.

\subsection{Estimated pre-WD companion properties}
\label{sec:wd}
The evidence presented in \ref{sec:basic} is alone sufficient to classify the central star of Hen~2-39 as a barium star. It is worth restating that historically the classification is based only on the properties of the s-process enhanced red giant (Bidelman \& Keeenan 1951), long before it was proven that these stars are binaries with WD companions (e.g. McClure et al. 1990). Since the pre-WD components of A~70 (Miszalski et al. 2012) and WeBo~1 (Siegel et al. 2012) were detected in the UV, we would therefore also expect a hot pre-WD companion to the observed central star in Hen~2-39 to be present. Unfortunately, we cannot prove our suspicion at this time as no space-based UV observations are yet available of Hen~2-39 at its low Galactic latitude of $b=-4.2^\circ$. Indirect evidence of a pre-WD is, however, present in the form of a highly ionized nebular emission line spectrum. The spectrum shows He~II $\lambda$4686/H$\beta=0.69$ (Sect. \ref{sec:neb}) and sufficient emission from low ionization species ([O~II], [N~II] and [S~II]) to suggest the nebula is optically thick. We can therefore estimate an Ambartsumyan or cross-over temperature for the ionising source using the method described in Kaler \& Jacoby (1989). Using their Eqn. 1 we find $T(\mathrm{cross})\sim225$ kK, consistent with an evolved pre-WD companion to the observed barium star. We can also estimate the observed visual magnitude of the pre-WD with Eqns. 2 and 3 of Kaler \& Jacoby (1989), the logarithmic extinction at H$\beta$ (Sect. \ref{sec:neb}) and the integrated H$\beta$ flux (Acker et al. 1992), to yield $V\sim22.5$ mag. We emphasise that these estimates are very uncertain. An improved estimate of these values was obtained as part of our \textsc{hoppla} analysis of the nebular spectrum, from which we find 195 kK for the Ambartsumyan temperature and $L=230$ $L_\odot$ for the pre-WD luminosity, assuming the Acker et al. (1992) H$\beta$ flux and a distance of 5.7 kpc (Sect. \ref{sec:distance}).

\section{Discussion}
\subsection{Distance and carbon star classification}
\label{sec:distance}
We estimated the distance to Hen~2-39 using the statistical Shklovsky distance scale of Stanghellini, Shaw \& Villaver (2008). This distance scale uses Magellanic Cloud PNe as calibrators to improve the Cahn, Kaler \& Stanghellini (1992) distance scale. An equivalent to the 5 Ghz radio flux was derived using the Acker et al. (1992) H$\beta$ flux [log $F(H\beta)=-12.4$ dex] and c(H$\beta$)=0.53 in Eqn. 6 of Cahn et al. (1992). With the nebular radius of 6.55\arcsec\ this gives a $\tau$ value of 4.56 and hence to a distance estimate of 5.7 kpc using Eqns. 1, 2 and 3b of Stanghellini et al. (2008). At this distance the absolute magnitudes of the nucleus are $M_V=+1.55$ and $M_K=-1.57$ mag. This is similar to $M_V=+1.30$ mag estimated for WeBo~1 based on the Cahn et al. (1992) distance scale (Bond et al. 2003). An additional comparison can be made using stars of similar spectral type to our R3 classification (Barnbaum et al. 1996), which nominally places Hen~2-39 amongst the peculiar group of early R-type stars (e.g. Dominy 1984; Wallerstein \& Knapp 1998; Zamora et al. 2009). Knapp, Pourbaix \& Jorissen (2001) used \emph{Hipparcos} parallaxes to determine the absolute magnitudes to several early R-type stars. The authors found an average of $M_V=0.81\pm1.03$ for the R3 class and many stars show similar $M_V$ and $M_K$ values to Hen~2-39 (e.g. HIP 19269 with $M_V=1.82$ and $M_K=-1.83$). The latter similarity gives us some assurance in the otherwise very uncertain distance estimate to Hen~2-39.  

Table \ref{tab:compare} shows several other properties that are in common with early R-type carbon stars. This raises the interesting question whether Hen~2-39 could be considered an early R-type carbon star. This class is peculiar because of the lack of evidence for binary mass transfer (McClure 1997) and s-process enhancement (Dominy 1984; Zamora et al. 2009). This has led some authors to suggest a binary merger scenario to explain the class (e.g. McClure 1997; Izzard et al. 2007; Angelou \& Lattanzio 2008) with differing results (Piersanti et al. 2010; Zhang \& Jeffery 2013). Hen~2-39, however, shows s-process enhancement and is most likely a binary system since a hot pre-WD should be present to ionise the nebula. This means Hen~2-39 most probably hosts a barium star nucleus instead of a canonical early R-type carbon star. Nevertheless, further study of Hen~2-39 may provide new insights into the difficult and unresolved problem surrounding the formation of early R-type carbon stars.

\begin{table*}
   \centering
   \caption{Comparison between properties of the visible nucleus of Hen~2-39 and early R stars.}
   \label{tab:compare}
   \begin{tabular}{llll}
      Property & Hen~2-39 & Early R & Reference\\
      \hline
      $T_\mathrm{eff}$ (K) & $4250\pm150$      & 4200--5000 & Dominy (1984); Wallerstein \& Knapp (1998)\\
                           &           & $>$3600 & Zamora et al. (2009)\\
      log g                & $2.0\pm0.5$       & 2.0 & Dominy (1984) \\
      $M_V$ (mag)           & 1.55   &  $0.81\pm1.03$  & Knapp et al. (2001)\\
                           &        &  0.4            & Dominy (1984)  \\
      $(V-K)_0$            & 3.12   & $<$4    & Knapp et al. (2001)\\
      s-process enhanced   & yes    & no      & Dominy (1984)        \\
      mass transfer binary & yes & no  & McClure (1997)\\
      {}[C/H]              & $0.42\pm0.02$ & $0.48\pm0.07$ & Dominy (1984); Wallerstein \& Knapp (1998)\\
      height above plane (pc)    & $-420$        & 300 & Knapp et al. (2001)\\

      \hline
   \end{tabular}
\end{table*}

Assuming an expansion velocity of 20 km s$^{-1}$ the kinematic ages at 5.7 kpc for the inner nebula and outer halo are $\sim$9000 yrs and $\sim$17000 yrs, respectively, with corresponding intrinsic radii of 0.18 and 0.34 pc. This configuration may be similar to the double shell found in LoTr1 (Tyndall et al. 2013). The 5.7 kpc distance also allows us to estimate the stellar radius of the nucleus. Assuming a bolometric correction of $-$0.8 mag for a K1 giant (Kaler 2011), combined with our derived $T_\mathrm{eff}=4250\pm150$ K, we find a radius of $R\sim12$ $R_\odot$. This corresponds to a rotation velocity of 111 km s$^{-1}$ for our preferred rotation period of 5.46 d, out of a maximum 218 km s$^{-1}$ (assuming $M=1.5$ $M_\odot$, a typical value for barium stars, Jorissen et al. 1998). If as the nebula suggests the inclination to the line of sight is low, e.g. $i=10^\circ$, we have $v \sin i=19$ km s$^{-1}$, explaining why we don't see broadened lines in our spectra. 
 
\subsection{AGB star models}
\label{sec:agb}
Heavy-element abundance predictions from low-mass AGB star models of 1.8 $M_{\odot}$,
$Z=0.01$ and 3.0 $M_{\odot}$, $Z=0.01$, were compared to the observations of Hen~2-39. 
We note that stellar evolutionary sequences of low metallicity AGB stars show deeper third dredge-up (TDU, e.g. Karakas et al. 2002), which leads to stronger carbon and s-process enrichment. Furthermore, s-process calculations of AGB stars also show a strong dependence on the initial metallicity (e.g. Busso et al. 2001). Nevertheless, the metallicity of these models ([Fe/H]$\sim$$-0.15$) is, within the errors of the nebular abundances (at least 0.1 dex for oxygen and worse for other elements), suitable for the Solar metallicity of Hen~2-39. Compared to Hen~2-39, the models have an argon abundance lower by 0.07 dex and oxygen abundances lower by 0.18-0.22 dex. The model masses selected are representative of the expected progenitor range that is poorly constrained by the non-Type I nebular abundances of Hen~2-39 (Sect. \ref{sec:chem}). The 3.0 M$_\odot$ model does not experience any hot bottom burning and produces a final composition consistent with non-Type I PNe. Most PNe likely originate from relatively low-mass progenitors as we use here with final core masses of $\sim$0.6 $M_{\odot}$ (e.g. Ferrario et al. 2005; Catal{\'a}n et al. 2008).

The 1.8 $M_{\odot}$, $Z=0.01$ model was previously described in Karakas, Campbell \& Stancliffe (2010), where we use the model with the most efficient convective overshoot (the $N_\mathrm{ ov} = 3$ case). The 3.0 $M_{\odot}$, $Z=0.01$ model is calculated using the same code and input physics, with the exception that no convective overshoot was used on the AGB as the model experienced deep TDU (see Shingles \& Karakas 2013).

For each model, we feed the evolutionary
sequences into a post-processing nucleosynthesis code and calculate
the abundances throughout each model star, as a function of time,
for elements from hydrogen through to bismuth. We use the same
code and nuclear network as described in Lugaro et al. (2012), and include 
a partially mixed zone of protons in exactly the same manner. 
The partially mixed region of protons is required for the formation 
of a $^{13}$C pocket, which
is responsible for releasing neutrons in the He-intershell via the
$^{13}$C($\alpha$, n)$^{16}$O reaction (see e.g. Gallino et al. 1998; Goriely et al. 2000; Busso et al. 2001; Kamath et al. 2012; Lugaro et al. 2012). 
The mass of the proton profile is a free parameter which we set as a 
constant mass. We experimented with different sizes for the partially 
mixed zone from $M_\mathrm{ mix} = 1, 2, 6 \times 10^{-3} M_{\odot}$
in the 1.8 $M_{\odot}$ model and $M_\mathrm{ mix} = 1, 2 \times 10^{-3}
M_{\odot}$ in the 3.0 $M_{\odot}$ case.
A mass of $M_\mathrm{ mix} = 2 \times 10^{-3} M_{\odot}$ results in a
$^{13}$C pocket that is $\approx$ 1/10$^\mathrm{ th}$ of the mass of the
He-intershell in low-mass AGB models.  Initial abundances for
the post-processing calculations are from Asplund et al. (2009), where we
scale the Solar abundances to a global metallicity of $Z=0.01$ for both models.

While the mass of the $^{13}$C pocket is highly uncertain and one of the biggest modelling uncertainties for the s-process, the mass can be constrained using observations of carbon enhanced metal-poor stars (Izzard et al. 2009; Bisterzo et al. 2012; Lugaro et al 2012), the number ratio of barium stars relative to G and K giants (e.g. Izzard et al. 2010), planetary nebulae (Shingles \& Karakas 2013), and post-AGB stars (e.g. Bona{\v c}i{\'c} Marinovi{\'c} et al. 2007a; De Smedt et al. 2012). For example, Bona{\v c}i{\'c} Marinovi{\'c} et al. (2007b) were able to constrain the mass of the $^{13}$C pocket within a factor of 2 using population synthesis models of the low-metallicity Large Magellanic Cloud post-AGB star MACHO 47.2496.8.

The 1.8 $M_{\odot}$ model experienced 13 thermal pulses (TPs), had a final
core mass of 0.585 $M_{\odot}$ and a radius of 603 $R_\odot$. The 3.0 $M_{\odot}$ model in comparison
had 22 TPs, a final core mass of 0.68 $M_{\odot}$ and a radius of 690 $R_\odot$. 
The number of TPs and the efficiency of the TDU help determine
how much material is mixed into the convective envelope during the AGB
phase. The 3.0 $M_\odot$ model dredged up 0.10 $M_{\odot}$ of material in
comparison to the 1.8 $M_\odot$ which mixed up 0.041 $M_{\odot}$, an increase of 
about a factor of 2.5 as a consequence of experiencing more TPs. The duration of
the thermally pulsing AGB is 1.5 Myr for the 3.0 $M_{\odot}$ model, shorter than
the 1.8 $M_{\odot}$ model which has a TP-AGB lifetime of 1.8 Myr. In both
cases the TP-AGB lifetime is only a small fraction of the total
AGB lifetime (14.8 Myr and 10.0 Myr for 1.8 and 3.0 $M_{\odot}$ models, respectively).

\subsection{AGB model results and comparison with Hen~2-39}
\label{sec:agbresults}
Most of the mass is ejected in the last couple of TPs, namely, 0.49 M$_\odot$ and 1.8 M$_\odot$ for the 1.8 and 3.0 M$_\odot$ models, respectively. Therefore, to produce the high level of C and Ba pollution observed in Hen~2-39, we would most likely expect that the star has reached this last TP. Table \ref{tab:results} gives the initial and final abundances for carbon and barium in the AGB models. It is evident that the 3.0 M$_\odot$ model produces similar amounts of carbon and less barium than the 1.8 M$_\odot$ model. While the 3.0 M$_\odot$ model dredges up more material from the He-intershell over the AGB lifetime, it is diluted into a much larger envelope. This means that the final [Ba/Fe] ratio is lower than or similar to that in the 1.8 M$_\odot$ model. To explain the large enrichment seen in Hen~2-39, one needs the largest contamination. Thus the 1.8 M$_\odot$ models are preferred. In these models the larger the carbon pocket mass, the larger the s-process production, so the larger pocket mass is also preferred ($6\times10^{-3}$ M$_\odot$).

\begin{table}
   \centering
   \caption{Initial and final AGB star model abundances for carbon and barium.}
   \label{tab:results}
   \begin{tabular}{llllll}
      \hline
      Mass:        & 1.8 & 1.8 & 1.8 & 3.0 & 3.0 \\
      Pocket size: & 1.0 & 2.0 & 6.0 & 1.0 & 2.0\\
      ($\times10^{-3}$ M$_\odot$)  & & & & & \\
      \hline
      {}[C/H]$_i$        & $-$0.17&$-$0.17 &$-$0.17 & $-$0.17 &$-$0.17 \\  
      {}[C/H]$_f$        & 0.67& 0.66& 0.60 & 0.62&  0.60\\
      {}[Ba/Fe]$_i$      & 0.00& 0.00& 0.00& 0.00& 0.00\\
      {}[Ba/Fe]$_f$      & 1.30& 1.64& 2.08& 1.53& 1.65\\
      \hline
   \end{tabular}
\end{table}

So can we explain the Hen~2-39 pollution by such models and realistic mass transfer?
The pollution level in Hen~2-39 is [C/H]=0.42, and [Ba/Fe]=1.5. To achieve this level of enrichment we require [C/H] 2.6 times higher than in the original (Solar) material of the star, and a [Ba/Fe] enrichment of a factor 30.
Initially, we have a red giant with Solar composition and envelope mass $M_\mathrm{2,env}$ which accretes $\Delta M_2$ from its AGB companion. The latter is enriched in carbon over Solar abundance by a factor $f_\mathrm{1,C}$  and in barium by a factor $f_\mathrm{1,Ba}$. This material is then diluted into the red giant envelope whose final mass is $M_\mathrm{2,env}  + \Delta M_2$ and the final overabundance is $f_\mathrm{2,C}=2.6$ and $f_\mathrm{2,Ba}=30$. This gives

\[ f_\mathrm{2,X} (M_\mathrm{ 2,env}  + \Delta M_2) = M_\mathrm{ 2,env}  +  \Delta M_2 f_\mathrm{ 1,X}, \]
which can be rearranged as
\[ \alpha = \Delta M_2/M_\mathrm{ 2,env}= (f_\mathrm{ 2,X}-1)/(f_\mathrm{ 1,X} - f_\mathrm{ 2,X}),\]
where $X$ is C or Ba and $\alpha$ is the amount of enrichment of $X$ required. 
Taking the 1.8 M$_\odot$ model, where at the final TP carbon is increased by 0.77 dex, i.e. $f_\mathrm{ 1,C}$ = 5.89,  we would therefore need to have $\alpha =  \Delta M_2 / M_\mathrm{ 2,env} = 0.48$. In order to reproduce the observed value of $f_\mathrm{ 2,Ba}=30$, such a value of $\alpha$ would then impose $f_\mathrm{ 1,Ba}$ $\sim$90. This is easily reproduced by the final amount of barium produced by the 1.8 M$_\odot$ model with a pocket size of $6\times10^{-3}$ M$_\odot$. 

If we were to take the 3.0 M$_\odot$ model with a pocket size of $2.0\times10^{-3}$ M$_\odot$ instead, its identical level of carbon enrichment implies the same value of $\alpha$ as the 1.8 M$_\odot$ model. Therefore, to reproduce the observed $f_\mathrm{2,Ba}=30$ we would also require $f_\mathrm{ 1,Ba}$ $\sim$90, however the 3.0 M$_\odot$ model only produces about half this amount. This significant difference demonstrates that reproducing Hen~2-39 would be more straightforward with lower mass AGB models.

We have made experiments following binary systems of various masses using the Binary Star Evolution code (Hurley, Tout \& Pols 2002). If the AGB star mass is assumed to be in the 1.8--2.0 M$_\odot$ range, then we conclude that the present red giant must have had an initial mass which is smaller by about 20\% that of the initial mass of the AGB star, after having accreted some material from its companion, in order for it to currently be a red giant. So we assume $M_\mathrm{2,0} = 1.45$ M$_\odot$. Such a star has a convective envelope of about 1.0 M$_\odot$, so this means that it must have accreted $\alpha$ times 1.0 $M_\odot$ = 0.48 $M_\odot$ from the AGB during the last phase.  This is an upper limit, as the AGB star lost smaller amounts of mass prior to the last TP that were also enriched to a lesser degree in C and s-process elements, and the giant was thus already partially contaminated. As the AGB star lost 0.49 M$_\odot$ during the last TP according to the model, it is thus possible for the giant to have accreted 0.48 M$_\odot$, provided the mass transfer was quasi-conservative. Such a quasi-conservative mass transfer could be the result of stable RLOF if the primary fills its Roche lobe just at the tip of the AGB phase and during the last TP, such that all the remaining mass of the AGB star is transferred onto the companion. Alternatively, in the case where no formal RLOF happened, wind RLOF, which is also thought to be quasi-conservative, could be sufficient (Mohamed \& Podsiadlowski 2007, 2011). Further detailed simulations are required to address the full history of this interesting system. The 1.8 $M_\odot$ model or one similar to it with a large pocket size is therefore a reasonable explanation for the progenitor of the pre-WD component likely present in Hen~2-39.

Although we have shown above that a 3.0 M$_\odot$ AGB star would have more difficulty to reproduce the level of barium enhancement observed in Hen~2-39, we have also looked at the mass transfer expected from such a massive AGB star. Here, again, we would need to have an accreting companion with a final mass close to 3.0 M$_\odot$ to be now seen as a red giant while the former AGB star is going through the PN phase. The limit seems to be $M_2=2.85$ M$_\odot$, which we adopt in the following. This mass is the result of accreting $\Delta M_2$ to an existing $M_\mathrm{2,0}$ star, where as shown above, we need $\Delta M_2= \alpha M_\mathrm{2,env}$. As the mass of the envelope of the accretor can be assumed to be, roughly, $M_\mathrm{2,env} \sim M_\mathrm{2,0} - 0.4$, we are thus led to $M_\mathrm{2,0} + \alpha (M_\mathrm{2,0} - 0.4) = 2.85$, or $M_\mathrm{2,0} \sim 2.05~M_\odot$. The companion has then accreted about 0.8 M$_\odot$ of the material produced by the AGB star during the last TP. As a 3.0 M$_\odot$ AGB star looses 0.9 M$_\odot$ during this last TP -- and will still loose as much material before finishing up as a WD -- there is again, in principle, plenty of material to accrete from to explain the final pollution.

The level of pollution experienced by the AGB star in Hen~2-39 is consistent with the abundances observed in the polluted giant companion, although we do not claim this is a unique or final solution. A detailed study of the evolution of the binary system, whose orbital period and mass function is currently unknown, is beyond the scope of this paper.

\subsection{Combining nebular and stellar abundances}
\label{sec:comb}
The abundance analysis results for Hen~2-39 may be compared to those obtained in a similar fashion for A~70 (Miszalski et al. 2012). There are no published nebular abundances for other PNe with barium CSPNe, a difficult task given their low nebular surface brightnesses. Despite the very small sample of two, we can see where each fits in the nebular s-process studies of Sterling \& Dinerstein (2008) and Karakas et al. (2009). The [Ba/Fe] estimates for the Type-I PN A~70 of $\sim$0.5 dex and the non-Type I PN Hen~2-39 of 1.50$\pm$0.25 dex are consistent with the trend found in these studies, namely that Type-I PNe are s-process poor. An additional object that could be considered, but does not seem to fit the trend, is Me~1-1 which appears to be a non-Type I bipolar PN (Shen, Liu \& Danziger 2004) with a K1--2 bright giant that is not s-process enhanced ([Ba/Fe]=0.0 dex, Pereira et al. 2008). 

With a larger sample it may be possible to meaningfully address the enigma surrounding Type-I PNe. While it may be that some Type-I PNe are be formed from single intermediate mass AGB stars, stars with M$\ga$4.5 $M_\odot$ required for hot bottom burning at Solar metallicities (Karakas 2010) are short-lived and rare according to the initial mass function. Some may be formed via rapidly rotating stars of $\sim$2.5--3.0 M$_\odot$ (e.g. Lagarde et al. 2012), in which case their tendency for low s-process enhancement may be explained by rotation suppressing the s-process (Herwig et al. 2003). This may help explain the Type-I abundances of A~70 at the large height of $\sim$2 kpc below the Galactic disk (Miszalski et al. 2012). Binary stellar evolution provides one pathway to inducing rapid rotation via accretion, as observed in barium CSPNe, but there may be many other possibilities in binaries given the large parameter space of binary evolution.

Without UV spectroscopy we are limited to a nebular carbon abundance that is enhanced 0.79 dex relative to Solar. Further UV observations are therefore needed to say whether the stellar [C/H]=0.42$\pm$0.02 dex enhancement matches that seen in the nebula.

\section{Conclusions}
\label{sec:conclusion}
We have presented evidence that reveal a barium central star binary in Hen~2-39, only the fourth known bona-fide case of a barium star surrounded by a PN. Our main conclusions are as follows:
\begin{itemize}
   \item The observed nucleus of Hen~2-39 is a $V=16.5$ mag red giant with a cool atmosphere $T_\mathrm{eff}=4250\pm150$ K with enhanced carbon ([C/H]=$0.42\pm0.02$ dex) and the s-process element barium ([Ba/Fe]=$1.50\pm0.25$ dex). A spectral class of C-R3 C$_2$4 was determined according to the Barnbaum et al. (1996) carbon star atlas and no lithium $\lambda$6707.83 was detected. As observed in other barium CSPNe, low-amplitude photometric variability with a possible period of 5.46 d was detected, most likely due to a fast rotation period caused by spin-up after mass accretion. This is also supported by the detection of weak chromospheric Ca~II K emission.
   \item An AGB star model with an initial mass of 1.8 $M_\odot$ and a relatively large carbon pocket size of $6\times10^{-3}$ $M_\odot$ produced sufficient amounts of carbon and barium that could plausibly produce the observed properties of Hen~2-39. The larger mass 3.0 $M_\odot$ model may also work, however this may be easily hampered by the greater degree of dilution of s-process material into the larger stellar envelope. Nevertheless, a lower mass AGB star offers a more straightforward solution to the problem. More detailed modelling is certainly warranted when the critical parameters of orbital period and mass function are measured.
   \item The nucleus shares many properties with early R-type carbon stars, whose origin is poorly understood and may involve a binary merger, the s-process enhancement and highly probable binary nature of Hen~2-39 suggest it may just be a barium star overlapping the parameter space of these peculiar stars. 
   \item The nebula shows an apparent ring-like morphology as also seen in the other PNe with barium CSPNe, WeBo~1 and A~70. A distance estimate of 5.7 kpc was made based on the nebular properties. At this distance the $T_\mathrm{eff}\sim195$ kK pre-WD ionizing source of the nebula has an estimated luminosity of $L=230$ $L_\odot$. Future UV observations are expected to reveal this component as already proven for A~70 (Miszalski et al. 2012) and WeBo~1 (Siegel et al. 2012). 
   \item The nebular abundance pattern of Hen~2-39 is close to that of the Sun and the average non-Type I PN. While Hen~2-39 is not a Type-I PN but is s-process rich, A~70 is a Type-I PN but less s-process rich). This seems to support the tendency for Type-I PNe to show lower s-process abundances (Karakas et al. 2009). 
\end{itemize}

\section*{Acknowledgments}
Most of the observations reported in this paper were obtained with the Southern African Large Telescope (SALT). Other supporting observations were made at Paranal Observatory, La Silla Observatory, and the South African Astronomical Observatory (SAAO). This work was co-funded under the Marie Curie Actions of the European Commission (FP7-COFUND). AIK is grateful for the support from the Australian Research Council for a Future Fellowship (FT110100475) and the NCI National Facility at the ANU for computing time. This work was partially supported by Spanish MICINN within the program CONSOLIDER INGENIO 2010, under grant ``Molecular Astrophysics: The Herschel and ALMA Era, ASTROMOL'' (ref.: CSD2009-00038)

SYNPHOT is a product of the Space Telescope Science Institute, which is operated by AURA for NASA. This research has made use of SAOImage \textsc{ds9}, developed by Smithsonian Astrophysical Observatory, the SIMBAD data base operated at CDS, Strasbourg, France, and makes use of data products from the 2MASS survey, which is a joint project of the University of Massachusetts and the Infrared Processing and Analysis Centre/California Institute of Technology, funded by the National Aeronautical and Space Administration and the National Science Foundation.

\label{lastpage}


\begin{thebibliography}{99}

\bibitem[Abate et al.(2013)]{2013A&A...552A..26A} Abate, C., Pols, O.~R., Izzard, R.~G., Mohamed, S.~S., \& de Mink, S.~E.\ 2013, A\&A, 552, A26 


\bibitem[Abia et al.(1993)]{1993A&A...272..455A} Abia, C., Boffin, H.~M.~J., Isern, J., \& Rebolo, R.\ 1993, A\&A, 272, 455 



\bibitem[Abia et al.(2002)]{2002ApJ...579..817A} Abia, C., Dom{\'{\i}}nguez, I., Gallino, R., et al.\ 2002, ApJ, 579, 817 

\bibitem[Acker et al.(1991)]{1991A&AS...89..237A} Acker, A., Raytchev, B., Koeppen, J., \& Stenholm, B.\ 1991, A\&AS, 89, 237 

\bibitem[Acker et al.(1992)]{1992secg.book.....A} Acker, A., Marcout, J., Ochsenbein, F., Stenholm, B., \& Tylenda, R.\ 1992, Garching: European Southern Observatory


\bibitem[Angelou \& Lattanzio(2008)]{2008PASA...25..155A} Angelou, G., \& Lattanzio, J.\ 2008, PASA, 25, 155 


   
   
\bibitem[Appenzeller et al.(1998)]{1998Msngr..94....1A} Appenzeller, I., et al.\ 1998, The Messenger, 94, 1 
\bibitem[Asplund et al.(2009)]{2009ARA&A..47..481A} Asplund, M., Grevesse, N., Sauval, A.~J., \& Scott, P.\ 2009, ARA\&A, 47, 481 


\bibitem[Barnbaum et al.(1996)]{1996ApJS..105..419B} Barnbaum, C., Stone, R.~P.~S., \& Keenan, P.~C.\ 1996, ApJS, 105, 419 




\bibitem[Bertin \& Arnouts(1996)]{1996A&AS..117..393B} Bertin, E., \& Arnouts, S.\ 1996, A\&AS, 117, 393 

\bibitem[Bessell(1979)]{1979PASP...91..589B} Bessell, M.~S.\ 1979, PASP, 91, 589 
   
\bibitem[Bidelman \& Keenan(1951)]{1951ApJ...114..473B} Bidelman, W.~P., \& Keenan, P.~C.\ 1951, ApJ, 114, 473 

\bibitem[Bisterzo et al.(2012)]{2012MNRAS.422..849B} Bisterzo, S., Gallino, R., Straniero, O., Cristallo, S., K{\"a}ppeler, F.\ 2012, MNRAS, 422, 849 
   
\bibitem[Boffin \& Jorissen(1988)]{1988A&A...205..155B} Boffin, H.~M.~J., \& Jorissen, A.\ 1988, A\&A, 205, 155 

\bibitem[Bona{\v c}i{\'c} Marinovi{\'c} et al.(2007)]{2007A&A...469.1013B} Bona{\v c}i{\'c} Marinovi{\'c}, A., Izzard, R.~G., Lugaro, M., \& Pols, O.~R.\ 2007a, A\&A, 469, 1013 
\bibitem[Bona al.(2007)]{2007A&A...472L...1B} Bona{\v c}i{\'c} Marinovi{\'c}, A., Lugaro, M., Reyniers, M., \& van Winckel, H.\ 2007b, A\&A, 472, L1 



\bibitem[Bond \& Neff(1969)]{1969ApJ...158.1235B} Bond, H.~E., \& Neff, J.~S.\ 1969, ApJ, 158, 1235 
\bibitem[Bond et al.(1993)]{1993IAUS..155..397B} Bond, H.~E., Ciardullo, R., \& Meakes, M.~G.\ 1993, Planetary Nebulae, 155, 397 
\bibitem[Bond et al.(2003)]{2003AJ....125..260B} Bond, H.~E., Pollacco, D.~L., \& Webbink, R.~F.\ 2003, AJ, 125, 260 

\bibitem[Breger et al.(1993)]{1993A&A...271..482B} Breger, M., Stich, J., Garrido, R., et al.\ 1993, A\&A, 271, 482 
\bibitem[Buckley et al.(2006)]{2006SPIE.6267E..32B} Buckley, D.~A.~H., Swart, G.~P., \& Meiring, J.~G.\ 2006, SPIE, 6267, 32  
\bibitem[Burgh et al.(2003)]{2003SPIE.4841.1463B} Burgh, E.~B., Nordsieck, K.~H., Kobulnicky, H.~A., et al.\ 2003, SPIE, 4841, 1463 

\bibitem[Busso et al.(1999)]{1999ARA&A..37..239B} Busso, M., Gallino, R., \& Wasserburg, G.~J.\ 1999, ARA\&A, 37, 239 
\bibitem[Busso et al.(2001)]{2001ApJ...557..802B} Busso, M., Gallino, R., Lambert, D.~L., Travaglio, C., \& Smith, V.~V.\ 2001, ApJ, 557, 802 


\bibitem[Buzzoni et al.(1984)]{1984Msngr..38.9B} Buzzoni, B., Delabre, B., Dekker, H., et al.\ 1984, The Messenger, 38, 9 




\bibitem[Castelli \& Kurucz(2006)]{2006A&A...454..333C} Castelli, F., \& Kurucz, R.~L.\ 2006, A\&A, 454, 333 


\bibitem[Catal{\'a}n et al.(2008)]{2008MNRAS.387.1693C} Catal{\'a}n, S., Isern, J., Garc{\'{\i}}a-Berro, E., \& Ribas, I.\ 2008, MNRAS, 387, 1693 



   

\bibitem[Corradi et al.(2003)]{2003MNRAS.340..417C} Corradi, R.~L.~M., Sch{\"o}nberner, D., Steffen, M., \& Perinotto, M.\ 2003, MNRAS, 340, 417 
\bibitem[Corradi et al.(2011)]{2011MNRAS.410.1349C} Corradi, R.~L.~M., Sabin, L., Miszalski, B., et al.\ 2011, MNRAS, 410, 1349 
\bibitem[Crawford et al.(2010)]{2010SPIE.7737E..54C} Crawford, S.~M., Still, M., Schellart, P., et al.\ 2010, Proc. SPIE, 7737E, 54 

\bibitem[\protect\citeauthoryear{Davis}{1987}]{1987PASP...99.1105D} Davis S.~P., 1987, PASP, 99, 1105 

\bibitem[De Smedt et al.(2012)]{2012A&A...541A..67D} De Smedt, K., Van Winckel, H., Karakas, A.~I., et al.\ 2012, A\&A, 541, A67 
   

\bibitem[Dominy(1984)]{1984ApJS...55...27D} Dominy, J.~F.\ 1984, ApJS, 55, 27 
   
\bibitem[Epchtein et al.(1999)]{1999A&A...349..236E} Epchtein, N., et al.\ 1999, A\&A, 349, 236 

\bibitem[Ferrario et al.(2005)]{2005MNRAS.361.1131F} Ferrario, L., Wickramasinghe, D., Liebert, J., \& Williams, K.~A.\ 2005, MNRAS, 361, 1131 


\bibitem[Gallino et al.(1998)]{1998ApJ...497..388G} Gallino, R., Arlandini, C., Busso, M., et al.\ 1998, ApJ, 497, 388 



   
\bibitem[Girard et al.(2007)]{2007A&A...463..265G} Girard, P., K{\"o}ppen, J., \& Acker, A.\ 2007, A\&A, 463, 265 

\bibitem[Goriely \& Mowlavi(2000)]{2000A&A...362..599G} Goriely, S., \& Mowlavi, N.\ 2000, A\&A, 362, 599 


   

\bibitem[Graham et al.(2004)]{2004MNRAS.347.1370G} Graham, M.~F., Meaburn, J., L{\'o}pez, J.~A., Harman, D.~J., \& Holloway, A.~J.\ 2004, MNRAS, 347, 1370 
\bibitem[Gray et al.(2011)]{2011AJ....141..160G} Gray, R.~O., McGahee, C.~E., Griffin, R.~E.~M., \& Corbally, C.~J.\ 2011, AJ, 141, 160 


\bibitem[Hamuy et al.(1994)]{1994PASP..106..566H} Hamuy, M., Suntzeff, N.~B., Heathcote, S.~R., et al.\ 1994, PASP, 106, 566 

\bibitem[Han et al.(1995)]{1995MNRAS.277.1443H} Han, Z., Eggleton, P.~P., Podsiadlowski, P., \& Tout, C.~A.\ 1995, MNRAS, 277, 1443 



\bibitem[Henize(1967)]{1967ApJS...14..125H} Henize, K.~G.\ 1967, ApJS, 14, 125 
\bibitem[Henry \& Worthey(1999)]{1999PASP..111..919H} Henry, R.~B.~C., \& Worthey, G.\ 1999, PASP, 111, 919 


\bibitem[Herwig et al.(2003)]{2003ApJ...593.1056H} Herwig, F., Langer, N., \& Lugaro, M.\ 2003, ApJ, 593, 1056 
   
\bibitem[Herwig(2005)]{2005ARA&A..43..435H} Herwig, F.\ 2005, ARA\&A, 43, 435 


   
\bibitem[Howarth(1983)]{1983MNRAS.203..301H} Howarth, I.~D.\ 1983, MNRAS, 203, 301 

\bibitem[Hurley et al.(2002)]{2002MNRAS.329..897H} Hurley, J.~R., Tout, C.~A., \& Pols, O.~R.\ 2002, MNRAS, 329, 897 
   


\bibitem[Izzard et al.(2007)]{2007A&A...470..661I} Izzard, R.~G., Jeffery, C.~S., \& Lattanzio, J.\ 2007, A\&A, 470, 661 



\bibitem[Izzard et al.(2009)]{2009A&A...508.1359I} Izzard, R.~G., Glebbeek, E., Stancliffe, R.~J., \& Pols, O.~R.\ 2009, A\&A, 508, 1359 


\bibitem[Izzard et al.(2010)]{2010A&A...523A..10I} Izzard, R.~G., Dermine, T., \& Church, R.~P.\ 2010, A\&A, 523, A10 

   



   
\bibitem[Jasniewicz et al.(1996)]{1996A&A...307..200J} Jasniewicz, G., Thevenin, F., Monier, R., \& Skiff, B.~A.\ 1996, A\&A, 307, 200 



\bibitem[Jeffries \& Stevens(1996)]{1996MNRAS.279..180J} Jeffries, R.~D., \& Stevens, I.~R.\ 1996, MNRAS, 279, 180 
   


\bibitem[Jones(2011)]{2011PhDT........15J} Jones, D.\ 2011, Ph.D.~Thesis, University of Manchester 

\bibitem[Jones et al.(2011)]{2011apn5.confP.111J} Jones, D., Pollacco, D., Faedi, F., \& Lloyd, M.\ 2011, Asymmetric Planetary Nebulae 5 Conference, 111P 

\bibitem[Jones et al.(2012)]{2012MNRAS.420.2271J} Jones, D., Mitchell, D.~L., Lloyd, M., et al.\ 2012, MNRAS, 420, 2271 

\bibitem[\protect\citeauthoryear{Jorissen, Manfroid, \& Sterken}{1992}]{1992A&A...253..407J} Jorissen A., Manfroid J., Sterken C., 1992, A\&A, 253, 407 

\bibitem[\protect\citeauthoryear{Jorissen et al.}{1995}]{1995A&A...301..707J} Jorissen A., Hennen O., Mayor M., Bruch A., Sterken C., 1995, A\&A, 301, 707 

\bibitem[\protect\citeauthoryear{Jorissen et al.}{1998}]{1998A&A...332..877J} Jorissen A., Van Eck S., Mayor M., Udry S., 1998, A\&A, 332, 877 



   

\bibitem[Kaler \& Jacoby(1989)]{1989ApJ...345..871K} Kaler, J.~B., \& Jacoby, G.~H.\ 1989, ApJ, 345, 871 

\bibitem[Kaler(2011)]{2011stsp.book.....K} Kaler, J.~B.\ 2011, Stars and their Spectra, by James B.~Kaler, Cambridge, UK: Cambridge University Press, 2011

\bibitem[Kamath et al.(2012)]{2012ApJ...746...20K} Kamath, D., Karakas, A.~I., \& Wood, P.~R.\ 2012, ApJ, 746, 20 
   

\bibitem[Karakas et al.(2002)]{2002PASA...19..515K} Karakas, A.~I., Lattanzio, J.~C., \& Pols, O.~R.\ 2002, PASA, 19, 515 




\bibitem[Karakas et al.(2009)]{2009ApJ...690.1130K} Karakas, A.~I., van Raai, M.~A., Lugaro, M., Sterling, N.~C., \& Dinerstein, H.~L.\ 2009, ApJ, 690, 1130 


\bibitem[Karakas(2010)]{2010MNRAS.403.1413K} Karakas, A.~I.\ 2010, MNRAS, 403, 1413 


\bibitem[Karakas et al.(2010)]{2010ApJ...713..374K} Karakas, A.~I., Campbell, S.~W., \& Stancliffe, R.~J.\ 2010, ApJ, 713, 374 
\bibitem[Karakas \& Lugaro(2010)]{2010PASA...27..227K} Karakas, A.~I., \& Lugaro, M.\ 2010, PASA, 27, 227 




\bibitem[Kingsburgh \& Barlow(1994)]{1994MNRAS.271..257K} Kingsburgh, R.~L., \& Barlow, M.~J.\ 1994, MNRAS, 271, 257 

\bibitem[Knapp et al.(2001)]{2001A&A...371..222K} Knapp, G., Pourbaix, D., \& Jorissen, A.\ 2001, A\&A, 371, 222 


   
\bibitem[Kobulnicky et al.(2003)]{2003SPIE.4841.1634K} Kobulnicky, H.~A., Nordsieck, K.~H., Burgh, E.~B., et al.\ 2003, SPIE, 4841, 1634 
\bibitem[Koppen et al.(1991)]{1991A&A...248..197K} K\"oppen, J., Acker, A., \& Stenholm, B.\ 1991, A\&A, 248, 197 

\bibitem[Kurtz \& Mink(1998)]{1998PASP..110..934K} Kurtz, M.~J., \& Mink, D.~J.\ 1998, PASP, 110, 934 

\bibitem[Kuschnig et al.(1997)]{1997A&A...328..544K} Kuschnig, R., Weiss, W.~W., Gruber, R., Bely, P.~Y., \& Jenkner, H.\ 1997, A\&A, 328, 544 


\bibitem[Lagarde et al.(2012)]{2012A&A...543A.108L} Lagarde, N., Decressin, T., Charbonnel, C., et al.\ 2012, A\&A, 543, A108 



\bibitem[Lenz \& Breger(2004)]{2004IAUS..224..786L} Lenz, P., \& Breger, M.\ 2004, The A-Star Puzzle, 224, 786 


\bibitem[Liu et al.(2006)]{2006MNRAS.368.1959L} Liu, X.-W., Barlow, M.~J., Zhang, Y., Bastin, R.~J., \& Storey, P.~J.\ 2006, MNRAS, 368, 1959 

\bibitem[Lugaro et al.(2012)]{2012ApJ...747....2L} Lugaro, M., Karakas, A.~I., Stancliffe, R.~J., \& Rijs, C.\ 2012, ApJ, 747, 2 

\bibitem[McClure(1997)]{1997PASP..109..256M} McClure, R.~D.\ 1997, PASP, 109, 256 


   
\bibitem[McClure et al.(1980)]{1980ApJ...238L..35M} McClure, R.~D., Fletcher, J.~M., \& Nemec, J.~M.\ 1980, ApJL, 238, L35 

\bibitem[Miszalski et al.(2011)]{2011A&A...531A.158M} Miszalski, B., Jones, D., Rodr{\'{\i}}guez-Gil, P., et al.\ 2011, A\&A, 531, A158 

\bibitem[Miszalski et al.(2012)]{2012MNRAS.419...39M} Miszalski, B., Boffin, H.~M.~J., Frew, D.~J., et al.\ 2012, MNRAS, 419, 39 

\bibitem[\protect\citeauthoryear{Miszalski, Boffin, \& Corradi}{2013}]{2013MNRAS.428L..39M} Miszalski B., Boffin H.~M.~J., Corradi R.~L.~M., 2013, MNRAS, 428, L39 


\bibitem[Mohamed \& Podsiadlowski(2007)]{2007ASPC..372..397M} Mohamed, S., \& Podsiadlowski, P.\ 2007, 15th European Workshop on White Dwarfs, 372, 397 


\bibitem[Mohamed \& Podsiadlowski(2011)]{2011ASPC..445..355M} Mohamed, S., \& Podsiadlowski, P.\ 2011, Why Galaxies Care about AGB Stars II: Shining Examples and Common Inhabitants, 445, 355 

   


   
   

\bibitem[Montez et al.(2010)]{2010ApJ...721.1820M} Montez, R., Jr., De Marco, O., Kastner, J.~H., \& Chu, Y.-H.\ 2010, ApJ, 721, 1820 
   
\bibitem[O'Donoghue et al.(2006)]{2006MNRAS.372..151O} O'Donoghue, D., Buckley, D.~A.~H., Balona, L.~A., et al.\ 2006, MNRAS, 372, 151 


\bibitem[Peimbert \& Torres-Peimbert(1983)]{1983IAUS..103..233P} Peimbert, M., \& Torres-Peimbert, S.\ 1983, Planetary Nebulae, Vol. 103. Reidel Publishing, Dordrecht, p. 233 

\bibitem[Pereira et al.(2008)]{2008A&A...477..535P} Pereira, C.~B., Miranda, L.~F., Smith, V.~V., \& Cunha, K.\ 2008, A\&A, 477, 535 


   

\bibitem[Piersanti et al.(2010)]{2010A&A...522A..80P} Piersanti, L., Cabez{\'o}n, R.~M., Zamora, O., et al.\ 2010, A\&A, 522, A80 


   
\bibitem[Porter et al.(2007)]{2007ApJ...657..327P} Porter, R.~L., Ferland, G.~J., \& MacAdam, K.~B.\ 2007, ApJ, 657, 327 

\bibitem[Shen et al.(2004)]{2004A&A...422..563S} Shen, Z.-X., Liu, X.-W., \& Danziger, I.~J.\ 2004, A\&A, 422, 563 

\bibitem[Shingles \& Karakas(2013)]{2013MNRAS.431.2861S} Shingles, L.~J., \& Karakas, A.~I.\ 2013, MNRAS, 431, 2861 






\bibitem[Siegel et al.(2012)]{2012AJ....144...65S} Siegel, M.~H., Hoversten, E., Bond, H.~E., Stark, M., \& Breeveld, A.~A.\ 2012, AJ, 144, 65 
\bibitem[Skrutskie et al.(2006)]{2006AJ....131.1163S} Skrutskie, M.~F., et al.\ 2006, AJ, 131, 1163 

\bibitem[Snodgrass et al.(2008)]{2008Msngr.13218S} Snodgrass, C., Saviane, I., Monaco, L., \& Sinclaire, P.\ 2008, The Messenger, 132, 18 



\bibitem[Stanghellini et al.(2008)]{2008ApJ...689..194S} Stanghellini, L., Shaw, R.~A., \& Villaver, E.\ 2008, ApJ, 689, 194 
\bibitem[Sterling \& Dinerstein(2008)]{2008ApJS..174..158S} Sterling, N.~C., \& Dinerstein, H.~L.\ 2008, ApJS, 174, 158 (SD08)




\bibitem[Strassmeier et al.(1997)]{1997A&A...322..511S} Strassmeier, K.~G., Hubl, B., \& Rice, J.~B.\ 1997, A\&A, 322, 511 



\bibitem[Theuns et al.(1996)]{1996MNRAS.280.1264T} Theuns, T., Boffin, H.~M.~J., \& Jorissen, A.\ 1996, MNRAS, 280, 1264 
   
\bibitem[Thevenin \& Jasniewicz(1997)]{1997A&A...320..913T} Thevenin, F., \& Jasniewicz, G.\ 1997, A\&A, 320, 913 

\bibitem[Tylenda et al.(1991)]{1991A&AS...89...77T} Tylenda, R., Acker, A., Raytchev, B., Stenholm, B., \& Gleizes, F.\ 1991, A\&AS, 89, 77 

\bibitem[Tyndall]{T2013} Tyndall, A. A., Jones, D., Boffin, H.M.J., Miszalski, B., Faedi, F., Lloyd, M., L\'opez, J.A., Martell, S., Pollacco, D., \& Santander-Garc\'ia M.\ 2013, MNRAS, in press 

\bibitem[van Dokkum(2001)]{2001PASP..113.1420V} van Dokkum, P.~G.\ 2001, PASP, 113, 1420 

\bibitem[Wallerstein \& Knapp(1998)]{1998ARA&A..36..369W} Wallerstein, G., \& Knapp, G.~R.\ 1998, ARA\&A, 36, 369 

\bibitem[Zamora et al.(2009)]{2009A&A...508..909Z} Zamora, O., Abia, C., Plez, B., Dom{\'{\i}}nguez, I., \& Cristallo, S.\ 2009, A\&A, 508, 909 


\bibitem[Zhang \& Jeffery(2013)]{2013MNRAS.430.2113Z} Zhang, X., \& Jeffery, C.~S.\ 2013, MNRAS, 430, 2113 


\end{thebibliography}
\end{document}